\documentclass[lettersize,journal]{IEEEtran}
\usepackage{amsmath,amsfonts}
\usepackage{amsthm}
\usepackage{array}
\usepackage[caption=false,font=normalsize,labelfont=sf,textfont=sf]{subfig}
\usepackage{textcomp}
\usepackage{stfloats}
\usepackage{url}
\usepackage{verbatim}
\usepackage{graphicx}
\usepackage{cite}
\usepackage{caption}
\usepackage{subcaption}
\usepackage{xcolor}
\makeatletter
\newif\if@restonecol
\makeatother

\usepackage[linesnumbered,ruled,vlined]{algorithm2e}%[ruled,vlined]{
\usepackage{algpseudocode}
\begin{document}

\title{Minimizing Maximum Latency of Task Offloading for Multi-UAV-assisted
Maritime Search and Rescue}

\author{
Shuang~Qi,~\IEEEmembership{Student Member,~IEEE,}
Bin~Lin,~\IEEEmembership{Senior Member,~IEEE,}
Yiqin~Deng,~\IEEEmembership{Member,~IEEE,} 
Xianhao~Chen,~\IEEEmembership{Member,~IEEE,} 
and~Yuguang~Fang,~\IEEEmembership{Fellow,~IEEE}
\thanks{The work of Bin Lin was supported in part by the National Natural Science Foundation of China (No. 62371085, No. 51939001) and in part by the Fundamental Research Funds for the Central Universities (No. 3132023514). The work of Yiqin Deng was supported in part by the National Natural Science Foundation of China (No. 62301300). The work of Yuguang Fang was supported in part by the Hong Kong SAR Government under the Global STEM Professorship and the Hong Kong Jockey Club under JC STEM Lab of Smart City. \textit{(Corresponding author: Bin Lin.)}}
\thanks{Shuang Qi and Bin Lin are with the Information Science and Technology College, Dalian Maritime University, Dalian 116026, China (e-mail: qishuang\_0315@163.com; binlin@dlmu.edu.cn).}
\thanks{Yiqin Deng is with the School of Control Science and Engineering, Shandong University, Jinan, Shandong, 250061, China and also with the Shandong Key Laboratory of Wireless Communication Technologies, Jinan, Shandong, 250061, China (e-mail: yiqin.deng@email.sdu.edu.cn).}
\thanks{Xianhao Chen is with the Department of Electrical and Electronic Engineering and HKU Musketeers Foundation Institute of Data Science, University of Hong Kong, Pok Fu Lam, Hong Kong SAR, China (e-mail: xchen@eee.hku.hk).}
\thanks{Yuguang Fang is with the Department of Computer Science, City University of Hong Kong, Hong Kong, China (e-mail: my.fang@cityu.edu.hk).}
\thanks{This manuscript has been accepted by IEEE Transactions on Vehicular Technology, DOI: 10.1109/TVT.2024.3384570.}}
\maketitle

\begin{abstract}
Unmanned Aerial Vehicles (UAVs) play a crucial role in Maritime Search and Rescue (MSAR), contributing to the improvement of rescue efficiency and reduction of casualties. Typically, UAVs equipped with cameras collect data from disaster areas and transmit it to the shore-based rescue command centers. By deploying Mobile Edge Computing (MEC) servers, UAVs can pre-process video footage to reduce data transmission volume, thus reducing transmission delays. However, the limited computational capacity and energy of UAVs pose significant challenges to the efficiency of UAV-assisted MSAR systems. To address these problems, in this paper, we investigate a multi-UAV assisted MSAR system consisting of multiple Surveillance UAVs (S-UAVs) and a Relay UAV (R-UAV). Then, we formulate a joint optimization problem to minimize the maximum total latency among all S-UAVs via jointly making the computing offloading decisions, R-UAV deployment, and the association between a S-UAV and rescue targets while ensuring that all targets are monitored by S-UAVs. Since the formulated optimization problem is typically hard to solve due to its non-convexity, we propose an effective iterative algorithm by breaking it into three sub-problems. Numerical simulation results show the effectiveness of the proposed algorithm with various performance parameters.
\end{abstract}


\section{Introduction}
\IEEEPARstart{T}{he} increasing number of maritime activities and ships at sea have led to a significant increase in maritime accidents such as collisions and groundings. According to the Annual Overview of Maritime Casualties and Incidents in 2022 published by the European Maritime Safety Agency (EMSA), in the period from 2014 to 2021, the total number of reported fatalities and injuries in marine casualties and incidents is 563 and 6155, respectively \cite{ref1}. In order to mitigate human casualties, Maritime Search and Rescue (MSAR) operations have been conducted worldwide. These operations take place at sea, involving the search for distressed individuals and providing assistance to maritime authorities and coordination centers, such as coast guards \cite{ref2}. However, relying solely on manpower for search and rescue in the complex maritime environment and volatile weather conditions is highly inefficient and risky. Establishing a temporary MSAR system to provide shore-based Rescue Coordination Center (RCC) with real-time video information from a disaster area is an effective measure to determine more effective search and rescue strategies, thereby enhancing MSAR efficiency.

Traditional MSAR systems mainly consist of satellite communication networks and shore-based communication networks. For example, the Global Maritime Distress and Safety System (GMDSS) for maritime distress and rescue operations, safety, and routine communications was established by International Maritime Organization (IMO) in 1988 \cite{ref3}. However, for delay-sensitive tasks such as MSAR tasks, the long delay and limited bandwidth of the satellite communication and the limited distance of the shore-based communication pose serious challenges, especially for real-time video transmissions. 

Unmanned Aerial Vehicles (UAVs) are considered to provide significant advantages in public safety, search and rescue, and disaster management at sea due to their high flexibility \cite{ref4}. On the one hand, the lack of high-definition cameras capable of video recording for maritime missions makes it hard to acquire real-time video information from disaster areas. UAVs equipped with sensors and cameras can be deployed to track and monitor target areas \cite{ref5,ref6,ref7,ref8}. On the other hand, UAVs have the ability to establish temporary emergency communication systems, allowing for real-time transmissions of surveillance video and other data from a disaster area to a Base Station (BS) at shore, or providing communication services to users in that particular area \cite{ref9,ref10,ref11, ref12}. 

Minimizing transmission latency is crucial for improving the efficiency of video transmissions in MSAR. Due to the high volume of video data, one way is to reduce the transmission volumes for a specific MSAR mission, which attempts to make trade-off between transmission and computing. Thus, Mobile Edge Computing (MEC) is considered to be a promising technology to beef up the computational capabilities at the proximity of terminal devices with computation-demanding applications, such as video processing \cite{ref13, ref14}. By pre-processing video data at UAVs equipped with MEC server, the amount of data to be transmitted can be significantly reduced, thereby reducing transmission latency. Meanwhile, in MSAR system, each UAV is responsible for monitoring a part of the area. To develop an efficient and fair search and rescue strategy, RCC needs to assess the overall situation of the disaster area based on the surveillance footage of each UAV. Therefore, the objective is to minimize the maximum latency, so that the delay of surveillance videos among all S-UAVs should be as small as possible to assist RCC in making quick decisions. 

Based on the above observations, multiple rotary-wing UAVs installed with Edge Servers (ESs) are dispatched to a disaster area for a search and rescue mission. Multiple rotary-wing UAVs have different types. Compared to larger rotary-wing UAVs, smaller rotary-wing UAVs are more flexible but have less load capacities. For example, the H20 UAV developed by United Aircraft Corporation is a large rotary-wing UAV suitable for search and rescue. Therefore, in this paper, multiple small rotary-wing UAVs are used for monitoring and computing due to the advantage of flexible and a large rotary-wing UAV is used for relaying and computing due to the greater load capacities. Specifically, multiple rotary-wing UAVs equipped with cameras for video surveillance are responsible for monitoring drowning people in the sea disaster area. Such UAVs are called Surveillance UAVs (S-UAVs), which have limited payload capacity due to practical considerations. To overcome this limitation, we introduce an additional UAV with a larger payload, referred to as Relay UAV (R-UAV), to serve as a relay for data transmissions and enhance computational capabilities. Videos can be computed on S-UAVs or offloaded to the R-UAV to reduce data volume by removing redundant video frames, thereby reducing latency. In scenarios involving multiple UAVs for video transmissions, we aim to minimize the maximum latency among all S-UAVs, so that the surveillance video of each UAV is successfully transmitted with the least possible delay difference to improve decision-making efficiency. Therefore, how to reduce the processing latency for a multi-UAV assisted MSAR system by optimizing the positions of UAVs and task offloading is the focus of this paper. We summarize our contributions as follows.
\begin{itemize}
\item A multi-UAV assisted MSAR system is designed, which consists of a R-UAV and multiple S-UAVs. To capture the condition of drowning people more accurately, we adjust the positions of S-UAVs based on the association between the S-UAV and rescue targets. By ensuring monitoring integrity, it can observe drowning people at a closer distance and keep the drowning people in the center of the monitoring area.
\item We formulate a joint optimization problem for the computing offloading decisions, R-UAV deployment, and the association between the S-UAV and rescue targets, with the goal of minimizing the maximum total latency among all S-UAVs under the energy constraints of R-UAV and S-UAVs.
\item An iterative optimization algorithm is designed by decomposing the original problem into three tractable sub-problems to find the near optimal solution. Specifically, linearization and Successive Convex Approximation (SCA) are used to solve the offloading optimization sub-problem and R-UAV position optimization sub-problem, respectively. Branch and Bound (BnB) algorithm is used to solve the association optimization sub-problem.
\end{itemize}

The remainder of this paper is organized as follows. Section II reviews the related works. Section III introduces the MSAR architecture. In Section IV, we present the system model. In Section V, we present the problem formulation. In Sections VI and VII, the proposed algorithm and its evaluation are given, respectively. Finally, Section VII concludes this paper.

\section{Related Works}
In this section, we review the research efforts on computing offloading in MEC systems. Then, we overview the works related to this paper in the applications of UAVs in two scenarios, i.e., video transmission and emergency communication.
\subsection{Computing Offloading in MEC}
MEC was proposed, for computing offloading from the user perspective as it focuses on sending tasks to edge nodes (e.g., BSs) where ESs are deployed, allowing the edge nodes to assist users in processing the applications and relieve the burden from user equipments for computing and power conservation \cite{ref15}. Since the delay is one of the most important factors for MSAR tasks, we focus on the related works of MEC in reducing delay. 

Deng {\em et al.} \cite{ref16} presented a brief overview on computing offloading. Computing offloading includes binary offloading \cite{ref17,ref18,ref19} and partial offloading \cite{ref20,ref21,ref22}. Tang \emph{et al}. \cite{ref23} studied the computational task offloading with non-divisible and delay-sensitive tasks in a MEC system. Aiming to minimize the delay of tasks, the offloading decision was optimized. Yang \emph{et al}. \cite{ref24} considered a MEC system with multi-server multi-users, which formulated the optimization problem of joint offloading decision and computation resource allocation, with the goal of minimizing the total latency delay and energy consumption of mobile users. If the terminal equipment has strong computing abilities, partial offloading is an effective strategy to allow users to offload parts of tasks on MEC servers, to achieve lower latency than binary offloading. Zhao \emph{et al}. \cite{ref25} investigated a collaborative MEC system with multi-UAV where task offloading is addressed to minimize the sum of execution delays and energy consumption by jointly designing the UAVs’ position, task partition ratios, and transmit power. Chen \emph{et al}. \cite{ref26} considered delay-sensitive tasks offloading in emergency communication scenarios and proposed a latency minimization problem by jointly optimizing the offloading decision, offloading ratio, and UAV trajectory. Nonetheless, these works mainly focus on minimizing the MEC systems latency, while in this paper, we attempt to limit the video transmission delay and make the video transmission delay of each S-UAV as close as possible by minimizing the maximum latency, in order to reduce the difference in video transmission time with the hope that the jitter can be mitigated as well.

\begin{figure*}[ht]
\centering 
{\includegraphics[width=0.8\textwidth]{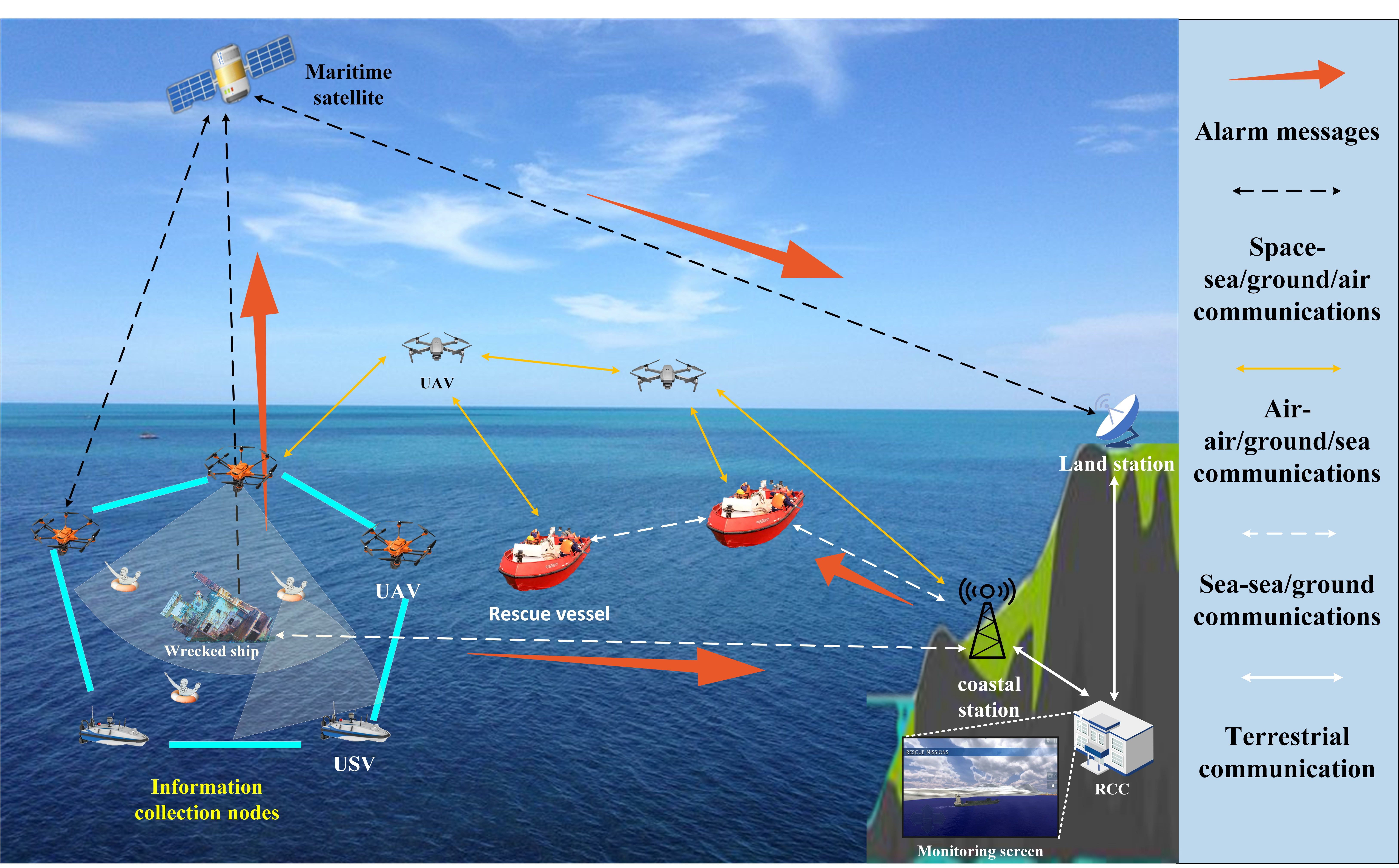}}
\caption{The MSAR procedure.}
\label{Fig.1}
\end{figure*}

\subsection{Application of UAVs in Video Transmissions}
UAVs can monitor specific areas through equipped high-definition cameras and deliver data over an ad-hoc network. Medeiros \emph{et al}. \cite{ref27} studied the path planning and deployment of multiple UAVs in a flood scenario to search for moving targets and deliver real-time video streams. Since communication resource is limited, a reasonable resource allocation scheme must be designed for effective video delivery to improve data transmission rate. Unfortunately, the video transmissions of UAVs impose significant pressure on wireless spectrum. 

By utilizing MEC technology and implementing video pre-processing, substantial savings in wireless spectrum can be achieved. Wang \emph{et al}. \cite{ref14} proposed a bandwidth-efficient architecture based on MEC to enable real-time video analysis of UAVs. They demonstrated how MEC can significantly reduce the bandwidth requirements for video analysis in UAVs without compromising the timeliness or accuracy of the results. Sun \emph{et al}. \cite{ref28} proposed a flexible cloud-edge collaborative scheduling strategy based on a UAV for mobile edge video analysis. They jointly optimized task assignments and the position of UAV to minimize the weighted sum of execution time and energy consumption. Deng \emph{et al}. \cite{ref29} developed an energy-efficient UAV-assisted target tracking system to offload video processing tasks from the UAV to edge nodes along its flight trajectory. By jointly optimizing the UAV's transmission power and edge node selection strategy, a trade-off between UAV transmission energy consumption and processing time is achieved. Li \emph{et al}. \cite{ref30} focused on edge-aided maritime UAV systems and investigated the impact of video frame resolution on the size of computational tasks and detection accuracy, as well as their effects on task latency and energy consumption. They optimized the UAV's transmission power, local CPU frequency, offloading ratio, and bandwidth allocation among UAVs. The UAV tracking system was modeled as an energy optimization problem with task latency constraints.

\subsection{Application of UAVs in Emergency Communications}
Emergency communications often require sending real-time videos or images to remote command centers. UAVs can provide an alternative solution for emergency coverage of users. Therefore, determining the optimal position or flight trajectory of UAVs with efficient wireless resource allocation to improve communication efficiency is a crucial issue. Do-Duy \emph{et al}. \cite{ref31} employed UAVs as flying BSs for real-time recovery and maintenance of network connectivity in disaster and post-disaster scenarios. They proposed a user clustering model with Quality-of-Service (QoS) constraints to form multiple distributed ground user clusters within the disaster area, guiding the real-time deployment of UAVs, aiming to maximize the energy efficiency. Hu \emph{et al}. \cite{ref9} formulated a joint design problem to optimize bandwidth, transmission power, and UAV altitude in a UAV-enabled uplink emergency communication scenario under the urban environment. Lin \emph{et al}. \cite{ref32} proposed an adaptive UAV deployment scheme to solve the coverage problem of UAV-aided ground communications. The objective is to optimize the location of UAVs to cover as many ground nodes as possible to reduce communication energy consumption.

By pre-processing video data on an MEC server equipped with a UAV, the amount of transmitted data can be significantly reduced, thus reducing transmission latency \cite{ref33}. Yang \emph{et al}. \cite{ref34} designed a seaborne region exploration architecture based on UAVs and joint grouping of Unmanned Surface Vehicles (USVs), and used MEC to complete the exploration mission in an accident area. To maximize the system throughput, the transmission power of UAV was optimized. However, the optimization of computation resources was not taken into consideration. Zhan \emph{et al}. \cite{ref35} considered a UAV-enabled MEC architecture, where IoT devices were at fixed locations to conduct certain sensing tasks with computation requirements. Aiming to minimize the completion time, the UAV trajectory, offloading scheduling and resource allocation were jointly optimized. All aforementioned works, though bearing some similarity in sensing or communications, have not touched upon the rescue missions, which is the target of our paper. 

\section{Maritime Search and Rescue Procedure}
MSAR aims to search and rescue the distressed objects caused by maritime accidents, such as fire, collision, explosion and abandonment. MSAR operations usually consist of two phases of maritime search and rescue \cite{ref36}. In maritime environment, only relying on human participation has the disadvantages of low efficiency and long response time. In addition, traditional search and rescue equipment (e.g., helicopter and rescue ships) lacks flexibility. Therefore, flexible, low-cost and intelligent unmanned systems, such as UAVs, USVs and Underwater Unmanned Vehicles (UUVs) \cite{ref37,ref38}, are applied to MASR operations to perform maritime disaster area search missions and establish the emergency communication to provide data transmissions with low latency and high reliability. These unmanned systems significantly improve the efficiency and security of MSAR operations. Fig. \ref{Fig.1} shows the MSAR procedure. 

Firstly, when a wrecked accident occurs, the Distress Alerts Messages (DAMs) can be transmitted to the RCC through the equipped GMDSS, including the identification of ships, the position of wrecked ship, time and other information that may be useful for search and rescue \cite{ref39}. If the distance between the wrecked ship and the coastal station exceeds 150 n miles, DAMs can be transmitted to RCC by maritime satellites. Otherwise, by using Very High Frequency (VHF) band and Medium Frequency (MF) band, the wrecked ship can transmit DAMs directly or through other ships to relay DAMs to the coastal station \cite{ref3}. The coastal station transmits DAMs to RCC by the terrestrial communication networks.

Then, RCC receives the alarm messages and forwards DAMs to other vessels in close proximity to the wrecked ship. Meanwhile, RCC evaluates the accident situation and formulates the search and rescue plan. Using the location of the wreckage as a reference point, unmanned systems are dispatched to the disaster area rapidly to search, monitor, and track the drowning objects. Additionally, unmanned systems equipped sensors and cameras can collect image and video data, and transmit to rescue ships and RCC. For long-distance transmissions, UAVs and USVs can form a multi-hop relay communication system to assist data transmissions. Therefore, the real-time videos can be shown on the monitoring screen in RCC. Furthermore, rescuers arrive at the disaster area by rescue ships and operate search and rescue missions with unmanned systems cooperatively. For instance, once UAVs find the drowning objects, they can throw a life buoy to the drowning objects and send the information (e.g., locations and physical conditions) for rescuers to conduct rescue operations.

In summary, the MSAR system, which integrates various communication resources to provide emergency assistance and necessary information exchange, provides a crucial guarantee for the efficiency of a search and rescue mission.

\begin{figure}[t]
\centering 
{\includegraphics[width=0.5\textwidth]{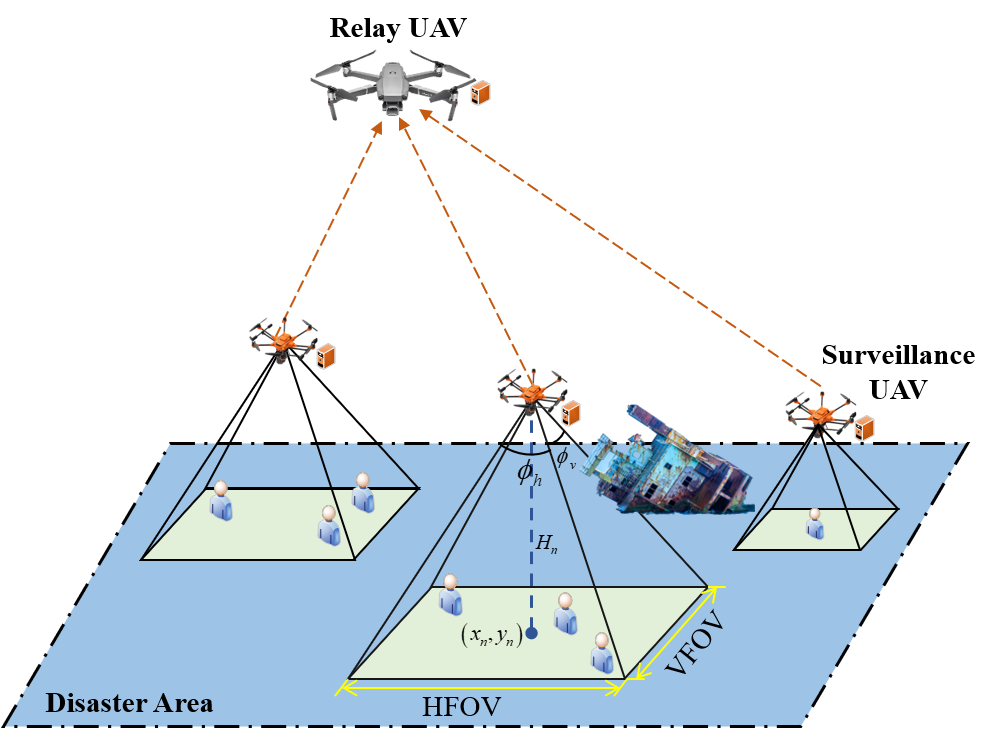}}
\caption{The multi-UAV assisted MSAR system.}
\label{Fig.2}
\end{figure}
\section{System Model}
\subsection{Network Model}
For illustrative purpose, we adopt a simplified model in this paper. As shown in Fig. \ref{Fig.2}, we present a multi-UAV assisted MSAR system, consisting of a R-UAV and $N$ S-UAVs. The set of S-UAVs is $\mathcal{N}=\{1,2,...,N\}$. Each UAV is equipped with an ES, and each S-UAV is equipped with a single antenna. R-UAV is equipped with two independent directional antennas for communication with BS and S-UAVs, respectively \cite{ref5}. Due to the high transmission power and reliable communication mechanism employed by R-UAV, we focus on video transmissions between S-UAVs and R-UAV. We assume that all S-UAVs transmit videos to R-UAV with Time Division Multiple Access (TDMA) scheme. We discretize the total search and rescue time $T$ into $K+1$ equal time slots, i.e., $k\in\mathcal{K}=\{0,1,2,...,K\}$, and each time slot is chosen to be small enough such that UAVs and targets can be considered to be stationary within each time slot. We denote the 3-D coordinates of R-UAV by $\boldsymbol{q}_M(k)=(x_M(k),y_M(k),H_M(k))$.

Assuming that each S-UAV arrives at the disaster area according to a pre-set flight route and conducts search at a pre-set hovering point, which means that S-UAV’s initial position is known, denote as $\boldsymbol{q}_n(0)=(x_n(0),y_n(0),H_n(0))$. The setting of the initial position needs to ensure that the entire disaster area is covered and overlapping areas are allowed.

The surveillance video is captured by a camera installed on S-UAV. For analysis simplicity, the Field-of-View (FOV) of a camera is assumed to be a rectangle, where the length $H_{\text {FOV}}$ of the rectangle is the Horizontal FOV (HFOV) and the width $V_{\text {FOV}}$ is the Vertical FOV (VFOV). The horizontal and vertical angles of a camera are denoted as $\phi_{h}$ and $\phi_{v}$, respectively. Therefore, the HFOV and VFOV at time slot $k$ can be expressed as \cite{ref40}:
\begin{equation}
H_{\text {FOV}}(k)=2H_{n}(k) \tan (\frac{\phi_h}{2}),
\end{equation}
\begin{equation}
V_{\text {FOV}}(k)=2H_{n}(k) \tan (\frac{\phi_v}{2}),
\end{equation}
then, the area covered by S-UAV $n$ is 
\begin{equation}
\begin{aligned}
&L_{n}(k) \times W_{n}(k) \\
=&[x_{n}(k)-H_{n}(k) \tan (\frac{\phi_h}{2}),x_{n}(k)+H_{n}(k) \tan (\frac{\phi_h}{2})] \\
&\times [y_{n}(k)-H_{n}(k) \tan (\frac{\phi_v}{2}),y_{n}(k)+H_{n}(k) \tan (\frac{\phi_v}{2})].
\end{aligned}
\end{equation}

Assuming that search targets are randomly distributed in the disaster area, the set of targets is represented as $\mathcal{I}=\{1, 2,...,i,...,I\}$, where $N\leq I$. The location of target $i$ at time slot $k$ is $\mathbf{q}_{i}(k)=(x_{i}(k),y_{i}(k),0)$. As we allow multiple S-UAVs monitoring ranges to overlap, the selection of S-UAVs for monitoring targets within the coverage of multiple S-UAVs will affect the positions of the S-UAVs. To this end, we introduce a binary variable $\alpha_{i,n}(k)\in \{0,1\}$ to indicate whether a target $i$ is monitored by S-UAV $n$ during time slot $k$. If target $i$ is monitored by S-UAV $n$, $\alpha_{i,n}(k)=1$; Otherwise, $\alpha_{i,n}(k)=0$, i.e.,
\begin{equation}
\alpha_{i,n}(k)=\left\{
        \begin{aligned}
	& 1 \quad \text{if target}\quad i \quad \text{is monitored by S-UAV} \quad n, \\
	& 0 \quad \text{otherwise}.\\
	\end{aligned}
	\right
         .
\end{equation}

If target $i$ is not within the coverage of S-UAV $n$, i.e., $(x_{i}(0),y_{i}(0)) \notin L_{n}(0) \times W_{n}(0)$, then $\alpha_{i,n}(0)=0$. Moreover, $\alpha_{i,n}(k)$ should meet the constraint as follows:
\begin{equation}
\sum_{n\in\mathcal{N}}{\alpha_{i,n}(k)} \geq 1,
\end{equation}
which means that every target should be monitored by at least one S-UAV. 

When search and rescue targets are identified in the monitoring area, and the association between S-UAV and a target is determined according to the initial position of S-UAV and targets. Then, S-UAVs need to adjust their position according to the location of the associated search and rescue targets in order to obtain a clearer video effect. Therefore, the location of S-UAV $n$ at time slot $k$ is denoted as $\mathbf{q}_{n}(k)=(x_{n}(k),y_{n}(k),H_{n}(k))$.

\begin{figure}[t]
\centering 
{\includegraphics[width=0.45\textwidth]{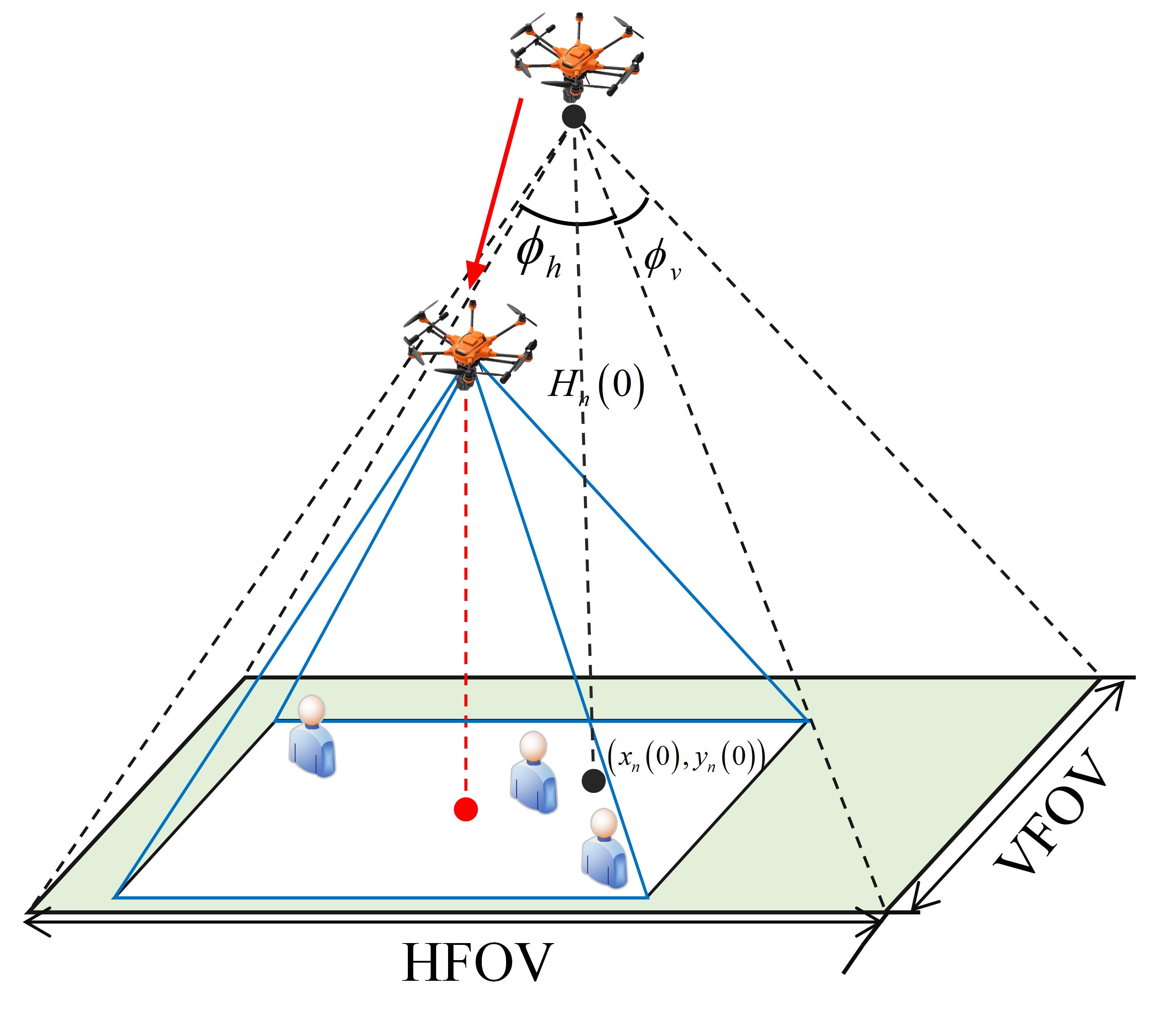}}
\caption{The illustration of the position of S-UAV adjustment process.}
\label{Fig.3}
\end{figure}

To ensure the distribution is relatively uniform within the monitoring area, we select the midpoint of the two points furthest apart in horizontal coordinates as the abscissa of the S-UAV. Similarly, the midpoint of the two points furthest apart in vertical coordinates as the ordinate of the S-UAV. Furthermore, to ensure that targets are always within the monitoring area of the S-UAV, the altitude of the S-UAV is adjusted depending on the locations of the two points furthest apart. Let $\mathcal{I}(n)=\{i|\alpha_{i,n}(k)=1, \forall i\in\mathcal{I}\}$ denote the set of targets that are monitored by S-UAV $n$. If $|\mathcal{I}(n)| \geq 2$, the horizontal coordinate of S-UAV $n$ can be expressed as
\begin{equation}
x_{n}(k)=\frac{\underset{i\in \mathcal{I}(n)}{\max}{\{x_{i}(k)\}}+\underset{i\in \mathcal{I}(n)}{\min}\{x_{i}(k)\}}{2},
\end{equation}
\begin{equation}
y_{n}(k)=\frac{\underset{i\in \mathcal{I}(n)}{\max}{\{y_{i}(k)\}}+\underset{i\in \mathcal{I}(n)}{\min}\{y_{i}(k)\}}{2},\\
\end{equation}
and the altitude of S-UAV $n$ is 
\begin{equation}
\begin{aligned}
H_{n}(k)=& \text{max}\{\frac{\underset{i\in \mathcal{I}(n)}{\max}{\{x_{i}(k)\}-\underset{i\in \mathcal{I}(n)}{\min}{\{x_{i}(k)\}}}}{2 \tan (\frac{\phi_h}{2})}, \\
&\frac{\underset{i\in \mathcal{I}(n)}{\max}{\{y_{i}(k)\}-\underset{i\in \mathcal{I}(n)}{\min}{\{y_{i}(k)\}}}}{2 \tan (\frac{\phi_v}{2})}\}+\gamma,
\end{aligned}
\end{equation}
where $\gamma$ is a constant to avoid targets falling on the boundary of the monitoring area. When $|\mathcal{I}(n)| =1$, the 3D-coordinate of S-UAV $n$ are $x_{n}(k)=x_{i\in\mathcal{I}(n)}(k), y_{n}(k)=y_{i\in\mathcal{I}(n)}(k), H_{n}(k)=\gamma$. When $|\mathcal{I}(n)|=0$, the 3D-coordinate of S-UAV $n$ is $\mathbf{q}_{n}(k)=(x_{n}(0),y_{n}(0),H_{n}(0))$, i.e., the location of S-UAV $n$ is not changed. Fig. \ref{Fig.3} illustrates the position of S-UAV adjustment process. The white rectangular area is the adjusted monitoring area of an S-UAV.

\subsection{Communication Model}
The communication link from an S-UAV to R-UAV is considered to be Line-of-Sight (LoS), and the channel attenuation mainly depends on the distance between S-UAV and R-UAV. Therefore, the channel gain between S-UAV $n$ and R-UAV $M$ can be obtained by the following formula (free-space model):
\begin{equation}
g_{n}(k)=\rho_{0}d_{n}^{-2}(k)=\frac{\rho_{0}}{{\Vert \mathbf{q}_{n}(k)-\mathbf{q}_{M}(k)\Vert}^2},
\end{equation}
where $d_{n}(k)$ denotes the distance between S-UAV $n$ and R-UAV at time slot $k$, and $\rho_{0}$ is the channel gain at the reference distance of $d_{0}=1$m. Let $\Gamma_{1}=\frac{\rho_{0}p_{n}(k)}{\sigma^2}$, then the transmission rate from S-UAV $n$ to R-UAV can be computed by the Shannon capacity for illustrative purpose: 
\begin{equation}
\begin{aligned}
R_{n}(k)& =B\log_{2}\Big(1+\frac{p_n(k)g_{n}(k)}{\sigma^2}\Big)\\
& =B\log_{2}\Big(1+\frac{\Gamma_{1}}{{\Vert \mathbf{q}_{n}(k)-\mathbf{q}_{M}(k)\Vert}^2}\Big),
\end{aligned}
\end{equation}
where $B$ is the bandwidth, $p_{n}(k)$ is the transmission power of S-UAV $n$ at time slot $k$, and $\sigma^2$ is the noise power at the receiver.

\subsection{Computing Model}
Define $\beta_{n}(k)\in\{0,1\}$ as the offloading indicator variable of S-UAV $n$ at time slot $k$, where $\beta_{n}(k)=1$ if S-UAV $n$ offloads a video processing task to R-UAV for computing at time slot $k$ and $\beta_{n}(k)=0$ if S-UAV $n$ executes the video task locally at time slot $k$, i.e.,
\begin{equation}
\beta_{n}(k)=\left\{
        \begin{aligned}
	& 1 \quad \text{if S-UAV } \quad n \quad \text{offloads task to R-UAV}, \\
        & 0 \quad \text{otherwise}.\\
	\end{aligned}
	\right
         .
\end{equation}
\paragraph{Computing on S-UAV} Let $f_{n}(k)$ and $S_{n}(k)$ denote the computation capability and the size of a video chunk (task) of S-UAV $n$ at time slot $k$, respectively. If there is no target in the coverage of S-UAV $n$ at time slot $k$, the video of S-UAV $n$ does not need to deliver, i.e., if $\sum_{i \in \mathcal{I}}{\alpha_{i,n}(K)=0}$, $S_{n}(k)=0$; Otherwise, if $\sum_{i \in \mathcal{I}}{\alpha_{i,n}(K)} \geq 1$, $S_{n}(k)\textgreater 0$. The size of the processed video chunk is $\hat{S}_{n}(k)=\mu_{n}S_{n}(k)$, and $\mu_{n}\in (0,1)$ is the ratio of the processed size and original size of a video chunk. Therefore, the local computing time is
\begin{equation}
t_{n}^{\text{loc}}=\frac{S_{n}(k)f_0}{f_{n}(k)},
\end{equation}
where $f_0$ is the computation resource cost to process per bit data. Then, the transmission latency of S-UAV $n$ to R-UAV after computing is 
\begin{equation}
t_{n,M}^{\text{loc}}=\frac{\mu_{n}S_{n}(k)}{B\log_{2}\Big(1+\frac{\Gamma_{1}}{{\Vert \mathbf{q}_{n}(k)-\mathbf{q}_{M}(k)\Vert}^2}\Big)}.
\end{equation}
\paragraph{Computing on R-UAV}
At time slot $k$, the offloading time from S-UAV $n$ to R-UAV is 
\begin{equation}
t_{n,M}^{\text{off}}=\frac{S_{n}(k)}{B\log_{2}\Big(1+\frac{\Gamma_{1}}{{\Vert \mathbf{q}_{n}(k)-\mathbf{q}_{M}(k)\Vert}^2}\Big)},
\end{equation}
and the computing time on R-UAV for S-UAV $n$'s video is 
\begin{equation}
t_{M}^{\text{off}}=\frac{S_{n}(k)f_0}{f_{M}(k)/{\underset{n\in \mathcal{N}}{\sum}{\beta_{n}(k)}}},
\end{equation}
where $f_{M}(k)$ is the CPU cycle frequency of R-UAV at time slot $k$, then $\frac{f_{M}(k)}{\sum_{n\in \mathcal{N}}{\beta_{n}(k)}}$ implies that R-UAV allocates computing resources to S-UAVs fairly.
\begin{table}
\renewcommand{\arraystretch}{1.3}
\caption{KEY NOTATIONS}
\label{table 1}
\centering
\begin{tabular}{|c|c|}
\hline
\textbf{Notations} & \textbf{Descriptions}\\
\hline
$\mathcal{N}$ & The set of S-UAVs\\
\hline
$\mathcal{I}$ & The set of rescue targets\\
\hline
$\alpha_{i, n}(k)$ & The indicator of a target monitored by S-UAV \\
\hline
$B$ & Signal bandwidth\\
\hline
$p_{n}(k)$ & The transmit power of S-UAV $n$ at time slot $k$ \\
\hline
$\beta_{n}(k)$ & The offloading decision of S-UAV $n$ at time slot $k$ \\
\hline
$S_{n}(k)$ & The data size of S-UAV $n$ at time slot $k$ \\
\hline
$f_0$ & The number of CPU cycles required for \\& computing 1-bit of data \\
\hline
$f_{n}(k)$ & The CPU frequency of S-UAV $n$ at time slot $k$\\
\hline
$f_{M}(k)$  & The CPU frequency of R-UAV at time slot $k$ \\
\hline
$E_{n}^{\text{hover}}(k)$ & The basic operational energy consumption \\& of S-UAV $n$ at time slot $k$ \\
\hline
$E_{M}^{\text{hover}}(k)$  & The basic operational energy consumption \\& of R-UAV at time slot $k$\\
\hline
\end{tabular}
\end{table}
In this paper, since we assume that each S-UAV needs to process the same amount of video data at the same time, we do not take into account the time required for transmission from R-UAV to BS. Therefore, the total latency of the S-UAV $n$ can be expressed as
\begin{equation}
T_{n}^{\text{total}}=(1-\beta_{n}(k))(t_{n}^{\text{loc}}+t_{n,M}^{\text{loc}})+\beta_{n}(k)(t_{n,M}^{\text{off}}+t_{M}^{\text{off}}).
\end{equation}
\subsection{UAV Energy Consumption Model}
In addition, due to the limited energy of UAVs, we consider the energy consumption constraints for both S-UAVs and R-UAV. In this section, we consider the basic operational energy consumption of UAVs (i.e., flying and hovering), as well as communication and computation energy consumption.
\paragraph{S-UAV energy consumption}
The communication energy consumption of S-UAV $n$ completes one transmission at time slot $k$ is
\begin{equation}
E_{n}^{\text{com}}(k)=p_{n}(k)[(1-\beta_{n}(k))t_{n,M}^{\text{loc}}+\beta_{n}(k)t_{n,M}^{\text{off}}].
\end{equation}

The computation energy consumption of S-UAV $n$ is calculated as 
\begin{equation}
E_{n}^{\text{comp}}(k)=(1-\beta_{n}(k))f_{n}^{2}(k)\zeta S_{n}(k)f_0,
\end{equation}
where $\zeta$ denotes the effective switched capacitance of CPU, and its value is relevant to the clip architecture \cite{ref41}.

Therefore, the total energy consumption of S-UAV $n$ is 
\begin{equation}
E_{n}(k)=E_{n}^{\text{com}}(k)+E_{n}^{\text{comp}}(k)+E_{n}^{\text{hover}}(k),
\end{equation}
where $E_{n}^{\text{hover}}(k)$ is the basic operational energy consumption of S-UAV $n$.

\paragraph{R-UAV energy consumption}
The computation energy consumption of R-UAV at time slot $k$ is given by
\begin{equation}
E_{M,n}^{\text{comp}}(k)=\sum_{n\in \mathcal{N}}{\beta_{n}(k)f_{M}^{2}(k)\zeta f_0S_{n}(k)}.
\end{equation}

Therefore, the total energy consumption of R-UAV at time slot $k$ is
\begin{equation}
E_{M}(k)=\sum_{n\in \mathcal{N}}\beta_{n}(k)E_{M,n}^{\text{comp}}(k)+E_{M}^{\text{hover}}(k),
\end{equation}
where $E_{M}^{\text{hover}}(k)$ is the basic operational energy consumption of R-UAV. The key notations of this paper and their descriptions are listed in TABLE ~\ref{table 1}.

\section{Problem Formulation}
In MSAR, ensuring the real-time surveillance of all distressed individuals by UAVs is of utmost importance. Additionally, to design efficient rescue strategies, the command personnel needs to have knowledge of the status of all drowning people. Therefore, the objective is to minimize the maximum total execution latency among all S-UAVs while satisfying the constraints of computing resources, energy consumption, and coverage requirements. We formulate a problem optimizing the computation offloading decision $\boldsymbol{\beta}=\{\beta_{n}(k), \forall n\in\mathcal{N}, k\in\mathcal{K}\}$, the location of R-UAV $\mathbf{q}_{M}=(x_{M}(k),y_{M}(k),H_{M}(k))$, and the association between S-UAV and target $\boldsymbol{\alpha}=\{\alpha_{i, n}(k), \forall n\in\mathcal{N},i\in\mathcal{I}, k\in\mathcal{K}\}$. The optimization problem {\bf P1} can be formulated as
\begin{alignat}{2}
\textbf{P1}\colon \mathop {{\rm{min}}}\limits_{_{\boldsymbol{\alpha}, \boldsymbol{\beta}, \mathbf{q}_{M}}} \quad & \max_{n\in\mathcal{N}}{T_{n}^{\text{total}}} \\
\mbox{s.t.}\quad
&\beta_{n}(k)\in\{0,1\}  \quad \forall n, k \tag{22a}\\
&\sum_{n\in\mathcal{N}}\beta_{n}(k)\leq N_0 \quad \forall k \tag{22b}\\
&\psi_{M}^{\text{l}} \leq \mathbf{q}_{M}(k)\leq \psi_{M}^{\text{u}} \tag{22c}\\
&\alpha_{i, n}(k)\in\{0,1\}  \quad \forall i, n, k \tag{22d}\\
&\sum_{n\in\mathcal{N}}\alpha_{i, n}(k) \geq 1 \quad \forall i, k \tag{22e}\\
&\alpha_{i, n}(k)=0  \quad \forall (x_i(0), y_i(0) \notin L_n(0)\times W_n(0) \tag{22f}\\
&E_{n}(k)\leq E_{n}^{\text{re}}(k) \quad \forall n, k \tag{22g}\\
&E_{M}(k)\leq E_{M}^{\text{re}}(k) \quad \forall k, \tag{22h}
\end{alignat} 
where constraint (22a) and (22b) show that offloading decision variable is a binary variable and R-UAV can execute the computation tasks of at most $N_0\leq N$ S-UAVs simultaneously. R-UAV is restricted in the considered area, as indicated by (22c). Constraint (22d) and (22e) show that the association variable is a binary variable and every target should be monitored by at least one S-UAV. Constraint (22f) indicates that when the target is not within the FOV of the S-UAV, $\alpha_{i, n}(k)=0$. Constraint (22g) and (22h) are the energy limitation, where $E_{n}^{\text{re}}(k)$ and $E_{M}^{\text{re}}(k)$ are the residual energy of S-UAV $n$ and R-UAV, respectively.

In \textbf{P1}, the offloading decision variable $\boldsymbol{\beta}$ and the association variable $\boldsymbol{\alpha}$ are binary, while the position of R-UAV $\mathbf{q}_{M}(k)$ is continuous. Furthermore, the positions of R-UAV and S-UAVs are coupled, where $\boldsymbol{\alpha}$ affects the positions of S-UAVs. When the position of S-UAV changes, the position of R-UAV and offloading decision are affected. These features make a Mixed-Integer Nonlinear Programming (MINLP) problem, which is typically hard to find the optimal solution. Thus, heuristic algorithm is needed to find suboptimal solution.

\section{Algorithmic Design}
In this section, we attempt to find approximate solutions to {\bf P1}. We reformulate the problem into a more tractable form by adding slack variables and decompose the reformulated problem into three sub-problems, namely, the offloading decision optimization, the position optimization of R-UAV, and the association optimization of S-UAVs and targets. We further develop an iterative algorithm to obtain the solution. Specifically, in the r-th iteration, we use the position of R-UAV and the association obtained in the last iteration to solve the offloading decision optimization problem, then use the known offloading decision and the association to solve the position optimization problem of R-UAV, and finally get the optimized association in a similar manner.
\subsection{Problem Reformulation}
By introducing a slack variable $s=\max \limits_{n\in\mathcal{N}}{T_{n}^{\text{total}}}$, \textbf{P1} can be re-written as \textbf{P2} \cite{ref42}, i.e., 
\begin{alignat}{2}
\textbf{P2}\colon \mathop {{\rm{min}}}\limits_{_{\boldsymbol{\beta},  \boldsymbol{\alpha}, \mathbf{q}_{M}},s} \quad & s \\
\mbox{s.t.}\quad
&(22\text{a})-(22\text{h}) \nonumber \\
&s\geq T_{n}^{\text{total}}, \forall n\in\mathcal{N}. \tag{23a}
\end{alignat}

Obviously, the problem is still a multi-variable non-convex problem. In the following, we propose an iterative algorithm to solve the problem \textbf{P2}. Specifically, problem \textbf{P2} is divided into three sub-problems. In the first sub-problem, the offloading decision is optimized with given R-UAV’s position $\mathbf{q}_{M}$ and the association $\boldsymbol{\alpha}$. In the second sub-problem, the R-UAV’s position is optimized with given offloading decision $\boldsymbol{\beta}$ and the association $\boldsymbol{\alpha}$. For any feasible $(\mathbf{q}_{M}, \boldsymbol{\beta})$, the third sub-problem is the association optimization problem. These sub-problems are solved in the following subsections. Afterward, we provide the overall iterative algorithm, and examine its convergence and complexity.
\subsection{Offloading Optimization}
In the first sub-problem {\bf SP1}, the offloading decision is optimized by giving the R-UAV’s position $\mathbf{q}_{M}$ and the association $\boldsymbol{\alpha}$. Then the offloading decision optimization sub-problem \textbf{SP1} can be expressed as follows:
\begin{alignat}{2}
\textbf{SP1}\colon \mathop {{\rm{min}}}\limits_{_{\boldsymbol{\beta},s}} \quad & s \\
\mbox{s.t.}\quad
&(22\text{a}),(22\text{b}), (22\text{g}), (22\text{h}), (23\text{a}). \nonumber
\end{alignat}

In \textbf{SP1}, due to the non-linearity of constraint (23a) with respect to variable $\boldsymbol{\beta}$, we define a variable $\xi_{n}=\beta_{n}(k)\sum_{n\in\mathcal{N}}\beta_{n}(k)$, then according to the constraints (22a) and (22b), $\xi_{n}$ should meet following constraints:
\begin{equation}
0 \leq \xi_{n} \leq N_0\beta_{n}(k),
\end{equation}
\begin{equation}
0 \leq \xi_{n} \leq \sum_{n\in\mathcal{N}}\beta_{n}(k),
\end{equation}
\begin{equation}
\xi_{n} \geq \sum_{n\in\mathcal{N}}\beta_{n}(k)-N_0(1-\beta_{n}(k)),
\end{equation}
thus, \textbf{SP1} is equivalent to \textbf{SP1-1}, i.e.,
\begin{alignat}{2}
\textbf{SP1-1}\colon  \mathop {{\rm{min}}}\limits_{_{\boldsymbol{\beta}, s, \boldsymbol{\xi}}} \quad & s  \\
\mbox{s.t.}\quad
&(22\text{a}),(22\text{b}), (22\text{g}) \nonumber \\
&(25), (26), (27) \nonumber \\
&s\geq \hat{T}_{n}^{\text{total}}, \forall n\in\mathcal{N}, \tag{28a} \\
&\sum_{n\in \mathcal{N}}\xi_{n}f_{M}^{2}(k)\zeta f_0S_{n}(k)+E_{M}^{\text{total}}(k)\leq E_{M}^{\text{re}}(k), \tag{28b}
\end{alignat}
where, 
\begin{equation}
\begin{aligned}
\hat{T}_{n}^{\text{total}}= & (1-\beta_{n}(k))(t_{n}^{\text{loc}}+t_{n,M}^{\text{loc}}) \\
 & +\beta_{n}(k)t_{n,M}^{\text{off}}+\xi_{n}\frac{S_{n}(k)f_0}{f_{M}(k)}.
\end{aligned}
\end{equation}

Obviously, equation (16) and equation (29) are equivalent. However, \textbf{SP1-1} is non-convex due to the binary variable $\beta_n(k)$. To make the problem more tractable, we relax the variable $\beta_n(k)$ to be a continuous variable, i.e., $0\leq \beta_n(k) \leq 1$. Therefore, \textbf{SP1-1} is transformed into \textbf{SP1-2},
\begin{alignat}{2}
\textbf{SP1-2}\colon \mathop {{\rm{min}}}\limits_{_{\boldsymbol{\beta}, s, \boldsymbol{\xi}}} \quad & s  \\
\mbox{s.t.}\quad
&0 \leq \beta_{n}(k) \leq 1 \quad \forall n\tag{30a} \\
&(22\text{b}), (22\text{g}) \nonumber \\
&(25), (26), (27), (28\text{a}),  (28\text{b}). \nonumber
\end{alignat}

This relaxation process means that the optimal value of problem \textbf{SP1-2} is a lower bound on the optimal value of problem \textbf{SP1-1}. It can be shown that \textbf{SP1-2} is a convex optimization problem, and then standard convex optimization tools such as CVX can be utilized to solve it efficiently.

\subsection{R-UAV's Position Optimization}
In the second sub-problem, the position of R-UAV is optimized with given offloading decision $\boldsymbol{\beta}$ and the association $\boldsymbol{\alpha}$. Then, the sub-problem \textbf{SP2} can be expressed as follows:
\begin{alignat}{2}
\textbf{SP2}\colon \mathop {{\rm{min}}}\limits_{{\mathbf{q}_{M}},s} \quad & s  \\
\mbox{s.t.}\quad
&(22\text{c}), (22\text{g}), (23), \nonumber 
\end{alignat}
since constraints (22g) and (23) are non-convex with respect to $\mathbf{q}_{M}(k)$, \textbf{SP2} is a non-convex problem. In the following, we introduce the slack variable $\lambda_{M}$ and transform \textbf{SP2} into problem \textbf{SP2-1}, 
\begin{alignat}{2}
\textbf{SP2-1}\colon \mathop {{\rm{min}}}\limits_{{\mathbf{q}_{M}},s, \lambda_{M}} \quad & s  \\
\mbox{s.t.}\quad
&(22\text{c}) \nonumber \\
&\frac{p_{n}(k)S_{n}(k)[(1-\beta_{n}(k))\mu+\beta_{n}(k)]}{\lambda_{M}} \nonumber \\
&\leq  E_{n}^{\text{re}}(k)-E_{n}^{\text{hover}}(k)-E_{n}^{\text{comp}}(k), \forall n \tag{32a} \\
&\frac{(\mu+\beta_{n}(k)(1-\mu))S_{n}(k)}{\lambda_{M}} \nonumber \\
&+[(1-\beta_{n}(k))t_{n}^{\text{loc}}+\beta_{n}(k)t_{M}^{\text{off}}] \nonumber \\
&\leq s, \forall n \tag{32b} \\
&B\log_{2}\Big(1+\frac{\Gamma_{1}}{{\Vert \mathbf{q}_{n}(k)-\mathbf{q}_{M}(k)\Vert}^2}\Big) \nonumber \\
& \geq \lambda_{M}, \forall n.  \tag{32c}
\end{alignat}

\emph{\textbf{Theorem 1}}: Problem \textbf{SP2-1} is equivalent to problem \textbf{SP2}.

\emph{\textbf{Proof}}: It can be shown that at the optimal solution to problem \textbf{SP2-1}, the equality of constraints in (32c) holds. Otherwise, the slack variable $\lambda_{M}$ can always be increased with other variables fixed, then the objective value of \textbf{SP2} keeps unchanged and all other constraints are still satisfied. Therefore, there always exists an optimal solution to \textbf{SP2-1} such that all constraints in (32c) are met with equality. Therefore, problem \textbf{SP2-1} is equivalent to problem \textbf{SP2}, which concludes the proof.      \qed

Obvirously, the problem \textbf{SP2-1} is still non-convex due to the non-convex constraints in (32c). To tackle this issue, the SCA technique can be applied, where the original function is approximated by a more tractable function at a given local point in each iteration \cite{ref35, ref42}. Note that although the (32c) is non-convex with respect to $\mathbf{q}_{M}(k)$, it is convex in ${\Vert \mathbf{q}_{n}(k)-\mathbf{q}_{M}(k)\Vert}^2$. Hence, we can consider norm squared as a single term. As a result, with given local point $\mathbf{q}_{M}^{r}(k)$, we obtain the following lower bound for the left-hand side in (32c) by applying the first-order Taylor expansion on logarithm function, i.e.,
\begin{equation}
\begin{aligned}
& \quad R_{n}(k) \\
&=B\log_{2}\Big(1+\frac{\Gamma_{1}}{{\Vert \mathbf{q}_{n}(k)-\mathbf{q}_{M}(k)\Vert}^2}\Big) \\
&\quad \geq B\log_{2}\Big(1+\frac{\Gamma_{1}}{{\Vert \mathbf{q}_{n}(k)-\mathbf{q}_{M}^{r}(k)\Vert}^2}\Big) \\
& \quad -\frac{B\Gamma_{1}\log_{2}e({\Vert \mathbf{q}_{n}(k)-\mathbf{q}_{M}(k)\Vert}^2-{\Vert \mathbf{q}_{n}(k)-\mathbf{q}_{M}^{r}(k)\Vert}^2)}{{\Vert \mathbf{q}_{n}(k)-\mathbf{q}_{M}^{r}(k)\Vert}^2({\Vert \mathbf{q}_{n}(k)-\mathbf{q}_{M}^{r}(k)\Vert}^2+\Gamma_{1})}\\
&=A_{n}^{r}(k)-I_{n}^{r}(k)({\Vert \mathbf{q}_{n}(k)-\mathbf{q}_{M}(k)\Vert}^2-{\Vert \mathbf{q}_{n}(k)-\mathbf{q}_{M}^{r}(k)\Vert}^2),  \\
&=\hat{R}_{n}(k)
\end{aligned}
\end{equation}
where, 
\begin{equation}
A_{n}^{r}(k)=B\log_{2}\Big(1+\frac{\Gamma_{1}}{{\Vert \mathbf{q}_{n}(k)-\mathbf{q}_{M}^{r}(k)\Vert}^2}\Big),   \nonumber
\end{equation}
\begin{equation}
I_{n}^{r}(k)=\frac{B\Gamma_{1}\log_{2}e}{{\Vert \mathbf{q}_{n}(k)-\mathbf{q}_{M}^{r}(k)\Vert}^2({\Vert \mathbf{q}_{n}(k)-\mathbf{q}_{M}^{r}(k)\Vert}^2+\Gamma_{1})}.  \nonumber
\end{equation}

At the local point $\mathbf{q}_{M}(k)=\mathbf{q}_{M}^{r}(k)$, the equality of (33) holds. With any given local point $\mathbf{q}_{M}^{r}$ and the lower bound expression in (33), problem \textbf{SP2-1} is approximated as follows:
\begin{alignat}{2}
\textbf{SP2-2}\colon \mathop {{\rm{min}}}\limits_{{\mathbf{q}_{M}},s, \lambda_{M}} \quad & s  \\
\mbox{s.t.}\quad
&(22\text{c}), (32\text{a}), (32\text{b}) \nonumber \\ 
&\hat{R}_{n}(k) \geq \lambda_{M}, \forall n\in\mathcal{N}.  \tag{34a}
\end{alignat}

Therefore, \textbf{SP2-2} is a standard convex optimization problem, which can be directly solved using CVX. The details of SCA algorithm is summarized in \textbf{Algorithm~\ref{alg:Algorithm 1}}. 
% 主体algorithm部分——start
    \begin{algorithm}[t]
        
        \caption{SCA for \textbf{SP2}}
        \label{alg:Algorithm 1}
         \LinesNumbered
        {
        \textbf{Initialization}: $\mathbf{q}_{M}^{r}(k)$, and let $r=0$
          
          \textbf{repeat}

           \quad Solve the convex problem SP2-2 for given local point $\mathbf{q}_{M}^{r}(k)$, and denote the optimal solution as $\mathbf{q}_{M}^{*}(k)$;

           \quad Update the local point $\mathbf{q}_{M}^{r+1}(k)=\mathbf{q}_{M}^{*}(k)$;

          \quad  Update $r=r+1$;
          }
        
        \textbf{Until} The minimum maximum latency $s$ converges within a prescribed accuracy.
        
    \end{algorithm}
% 主体algorithm部分——end

\subsection{Association Optimization of S-UAVs and Targets}
In the third sub-problem, the association of S-UAVs and targets is optimized for any feasible $(\mathbf{q}_{M},\boldsymbol{\beta})$, and thus, the sub-problem \textbf{SP3} can be expressed as follows:
\begin{alignat}{2}
\textbf{SP3}\colon\mathop {{\rm{min}}}\limits_{s, \boldsymbol{\alpha}} \quad & s  \\
\mbox{s.t.}\quad
&(22\text{d}), (22\text{e}), (22\text{f}), (22\text{g}), (23). \nonumber
\end{alignat}

Due to the 0-1 variable $\boldsymbol{\alpha}$, we use BnB algorithm to solve the sub-problem \textbf{SP3} directly.

    \begin{algorithm}
        \caption{Overall Iterative Algorithm for \textbf{P2}}
        \label{alg:Algorithm 2}
        \LinesNumbered

        {
          \textbf{Initialization}: $\mathbf{q}_{M}^{r}(k)$, $\boldsymbol{\alpha}^{r}$, and let $r=0$, the maximum number of iterations $r_{\text{max}}$;
          
          \textbf{repeat}

          \quad With given $\mathbf{q}_{M}^{r}(k)$ and $\boldsymbol{\alpha}^{r}$, obtain the optimal solution $\boldsymbol{\beta}^{r+1}$ by solving convex problem \textbf{SP1-2};

          \quad With given $\boldsymbol{\beta}^{r+1}$ and $\boldsymbol{\alpha}^{r}$, obtain the optimal solution $\mathbf{q}_{M}^{r+1}(k)$ by using \textbf{Algorithm 1};

          \quad For given $\boldsymbol{\beta}^{r+1}$ and $\mathbf{q}_{M}^{r+1}(k)$, obtain the optimal solution $\boldsymbol{\alpha}^{r+1}$ by using the BnB algorithm;

          \quad Update $r=r+1$;
          
          }
        
        \textbf{Until}  The minimum maximum latency $s$ of \textbf{P2} converges or the iteration time $r=r_{\text{max}}$.
    \end{algorithm}

\subsection{Overall Iterative Algorithm, Convergence, and Complexity}
Based on the results obtained in the previous three subsections, we design an iterative algorithm to solve \textbf{P2} for sub-optimal solutions, by applying SCA and alternating optimization. The details are summarized in \textbf{Algorithm 2}. The offloading decision, R-UAV locations and the association variable are optimized alternately in each iteration, by solving \textbf{SP1-2}, \textbf{SP2-2}, and \textbf{SP3}, correspondingly.

\emph{\textbf{Convergence Analysis}}: To prove the convergence of \textbf{Algorithm 2}, it is sufficient to prove that the objective value is non-increasing after the set of optimized variables is updated in each iteration. According to \textbf{Algorithm 2}, we have
\begin{equation}
\begin{aligned}
s^{r-1} & =s(\boldsymbol{\beta}^{r-1}, \mathbf{q}_{M}^{r-1}, \boldsymbol{\alpha}^{r-1}) \overset{\text{(a)}}{\geq} s(\boldsymbol{\beta}^{r}, \mathbf{q}_{M}^{r-1}, \boldsymbol{\alpha}^{r-1}) \\
& \overset{\text{(b)}}{\geq} s(\boldsymbol{\beta}^{r}, \mathbf{q}_{M}^{r}, \boldsymbol{\alpha}^{r-1}) \overset{\text{(c)}}{\geq} s(\boldsymbol{\beta}^{r}, \mathbf{q}_{M}^{r}, \boldsymbol{\alpha}^{r})=s^{r}.   
\end{aligned}
\end{equation}
Giving the position of R-UAV $\mathbf{q}_{M}^{r-1}$ and the association relationship $\boldsymbol{\alpha}^{r-1}$ of S-UAVs and targets, $\boldsymbol{\beta}^{r}$ is the sub-optimal solution to problem \textbf{SP1-2} in the $r$-th iteration, thus satisfying inequality (a). Similarly, with R-UAV's position optimization based SCA with given $\boldsymbol{\beta}^{r}$ and $\boldsymbol{\alpha}^{r-1}$, the inequality (b) holds apparently. For the variable $\boldsymbol{\alpha}^{r}$, BnB algorithm solves the sub-problem \textbf{SP3} with given $\boldsymbol{\beta}^{r}$ and $\mathbf{q}_{M}^{r}$. The lower bound of the problem can be obtained in each iteration, so that the inequality (c) holds. In conclusion, it can be observed that after optimizing the set of variables in each iteration, the objective value is non-increasing and lower bounded. Therefore, \textbf{Algorithm 2} is convergent.
 \begin{figure}
\centering 
{\includegraphics[width=0.5\textwidth]{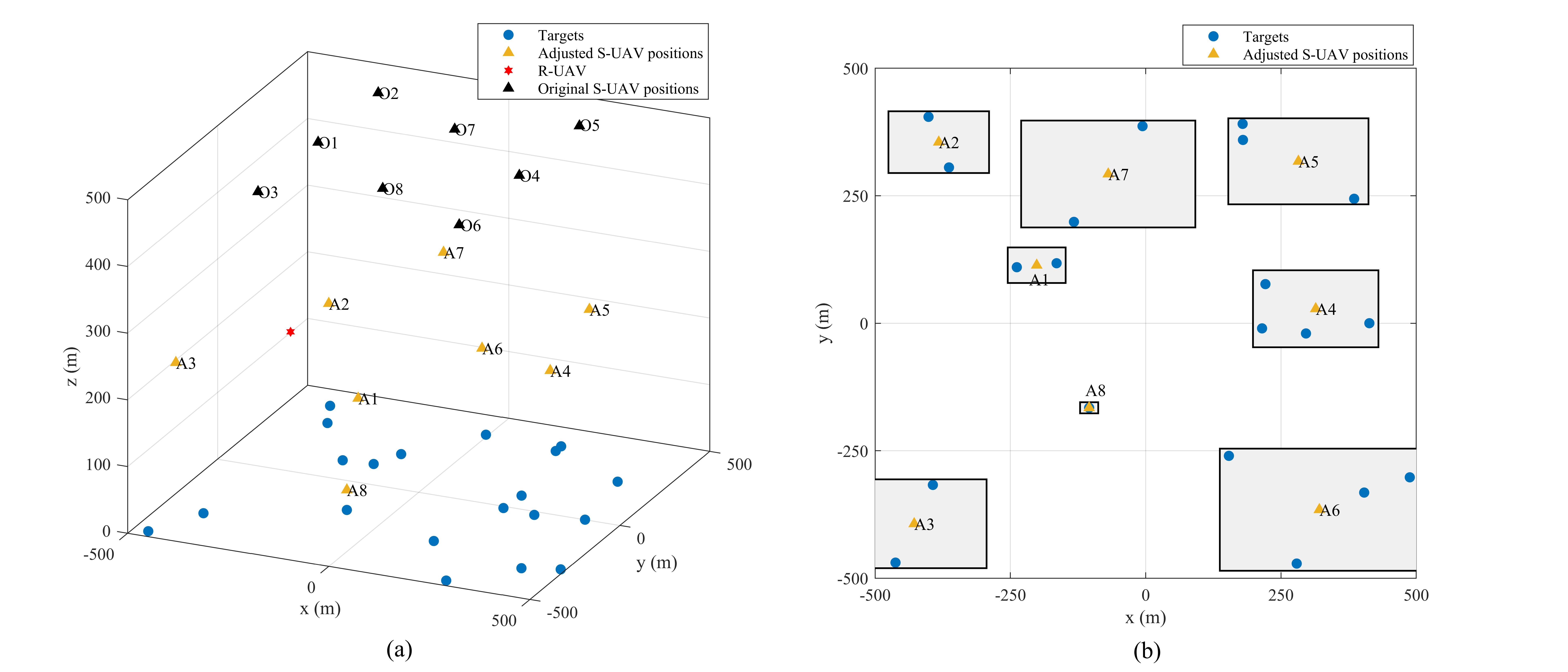}}
\caption{The distribution of S-UAVs, R-UAV and rescue targets. (a) locations of S-UAVs, R-UAV and rescue targets; (b) top view.}
\label{Fig.4}
\end{figure}

\emph{\textbf{Complexity Analysis}}: The complexity of the overall iterative algorithm includes the complexity of solving the problem \textbf{SP1-2} and \textbf{Algorithm 1}, and solving \textbf{SP3} with the BnB algorithm. Firstly, problem \textbf{SP1-2} is a standard convex optimization problem and can be solved by using the convex solver CVX with interior-point method, whose complexity are given by $\mathcal{O}((NK)^{3.5}\log(1/\epsilon))$ with given solution accuracy $\epsilon$ \cite{ref43}, since the number of optimization variables depends on $N$ and $K$. The complexity of problem \textbf{SP2} solved by SCA algorithm is $\mathcal{O}(L_{\text{S}}K^3)$, where $L_\text{S}$ is the number of iterations for \textbf{Algorithm 1} to converge. The complexity of BnB algorithm for solving the sub-problem \textbf{SP3} is $\mathcal{O}(2^{d}IN)$. Due to the binary variable $\boldsymbol{\alpha}$, then the number of branches per node is 2, the size of the problem is $IN$, and $d$ is the search depth. Therefore, the computational complexity of \textbf{Algorithm 2} is given as $\mathcal{O}(L_{\text{OIA}}((NK)^{3.5}\log(1/\epsilon)+L_{\text{s}}K^3+2^{d}IN))$, where $L_{\text{OIA}}$ is the number of iterations for \textbf{Algorithm 2} to converge.

\section{Performance Evaluation}
In this section, we conduct extensive experiments to evaluate the performance of our proposed algorithm.
\subsection{Parameter Settings}
The disaster sea area is 1 km$\times$1 km, which refers to the area that S-UAVs can monitor targets and capture videos. The number of S-UAVs and search and rescue targets are 8 and 20, respectively. The bandwidth is 10 MHz. The maximum transmit power of S-UAV is 1 W. The computational capabilities of S-UAV and R-UAV are set to $f_n=0.2$ GHz and $f_M=2$ GHz, respectively. Assuming that the surveillance video of each S-UAV is divided into multiple video chunks, and the size of each video chunk is randomly generated between [200,300] KB. Additionally, we set $\phi_h=58.4^{\circ}$, $N_0=4$, $p_n=0.8$ W, $\phi_v=40^{\circ}$, $f_0=1000$ cycles, $\rho_0=-60$ dB, $\sigma^2=-114$ dBm, $\gamma=30$ m, and $\zeta=10^{-28}$ \cite{ref26}\cite{ref30}.

The search and rescue targets are randomly distributed and S-UAVs are evenly distributed over the top of the monitoring area. Fig. \ref{Fig.4} shows the location distribution of S-UAVs, R-UAVs and rescue targets after an experiment. In  Fig. \ref{Fig.4}(a), the initial altitude of S-UAVs is set to 500 m. To capture video of drowning people, S-UAVs optimize positions. Accordingly, Fig. \ref{Fig.4}(b) shows the coverage relationship between S-UAVs and targets in a top view, where gray rectangles are the FOV of S-UAVs. It satisfies the requirement of covering all targets.
\begin{figure}[t]
\centering 
{\includegraphics[width=0.45\textwidth]{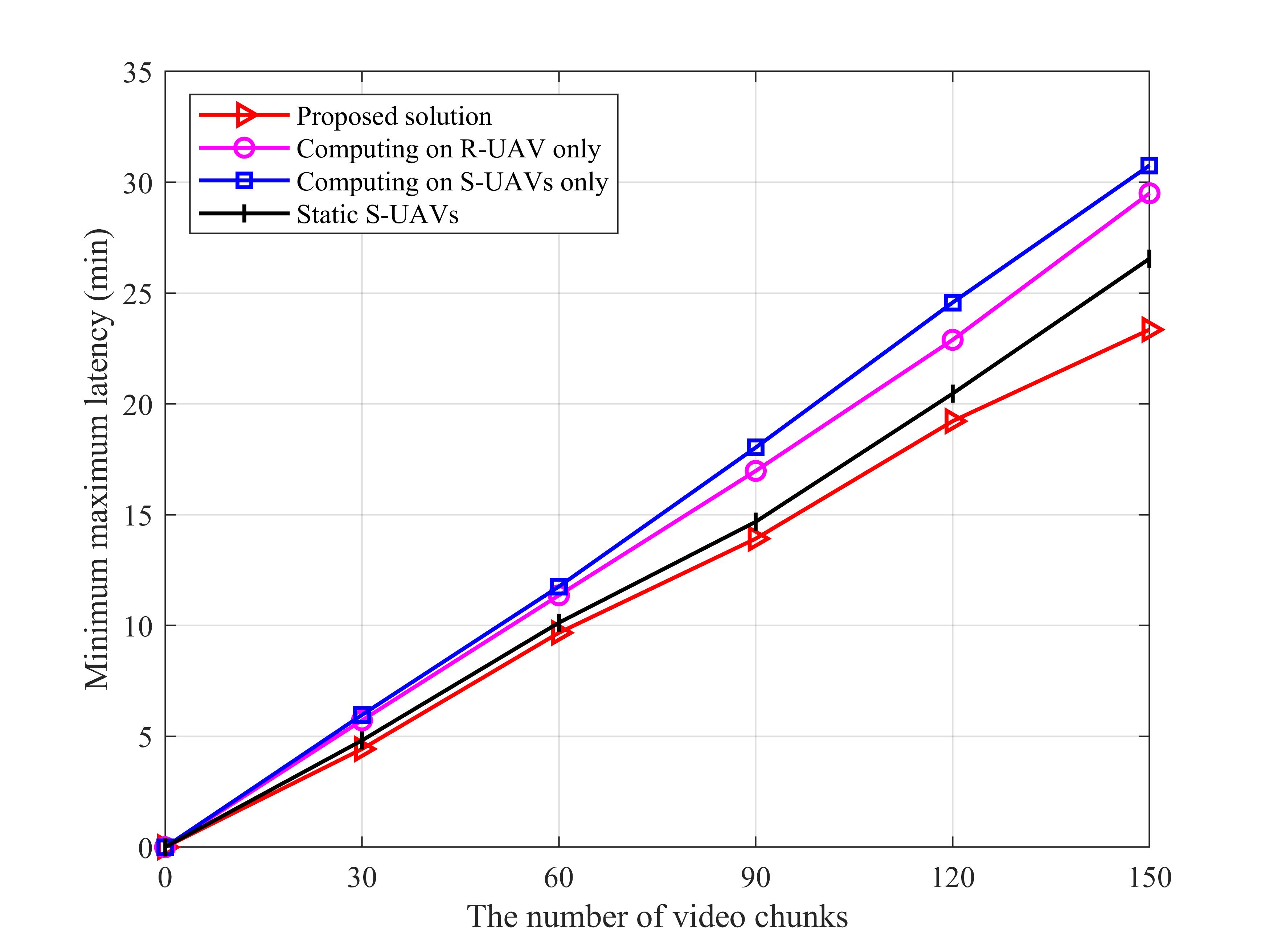}}
\caption{The minimum maximum latency versus the number of video chunks.}
\label{Fig.5}
\end{figure}

\begin{figure}[t]
\centering 
{\includegraphics[width=0.45\textwidth]{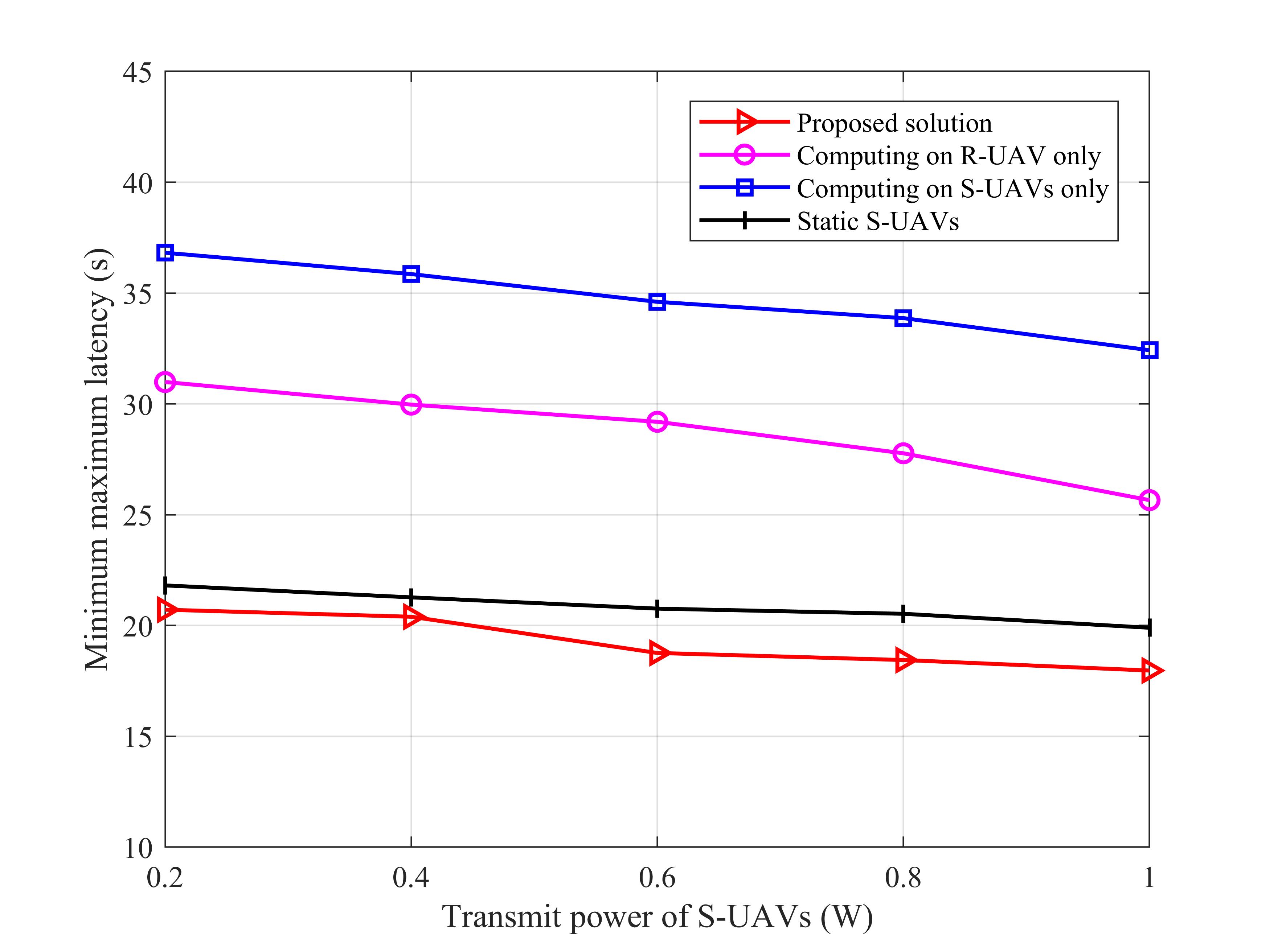}}
\caption{The minimum maximum latency versus the transmit power of S-UAVs.}
\label{Fig.6}
\end{figure}
\subsection{Numerical Results}
To demonstrate the performance improvement achieved by the proposed solution, we compare our proposed scheme with three benchmark schemes. These benchmark schemes include Computing on S-UAVs only, Computing on R-UAV only, and Static S-UAVs. The Static S-UAVs scheme means that when S-UAVs arrive at initial positions, regardless of the locations of rescue targets they are monitoring, S-UAVs remain in the initial positions and do not adjust their positions.

\begin{figure}[t]
\centering 
{\includegraphics[width=0.45\textwidth]{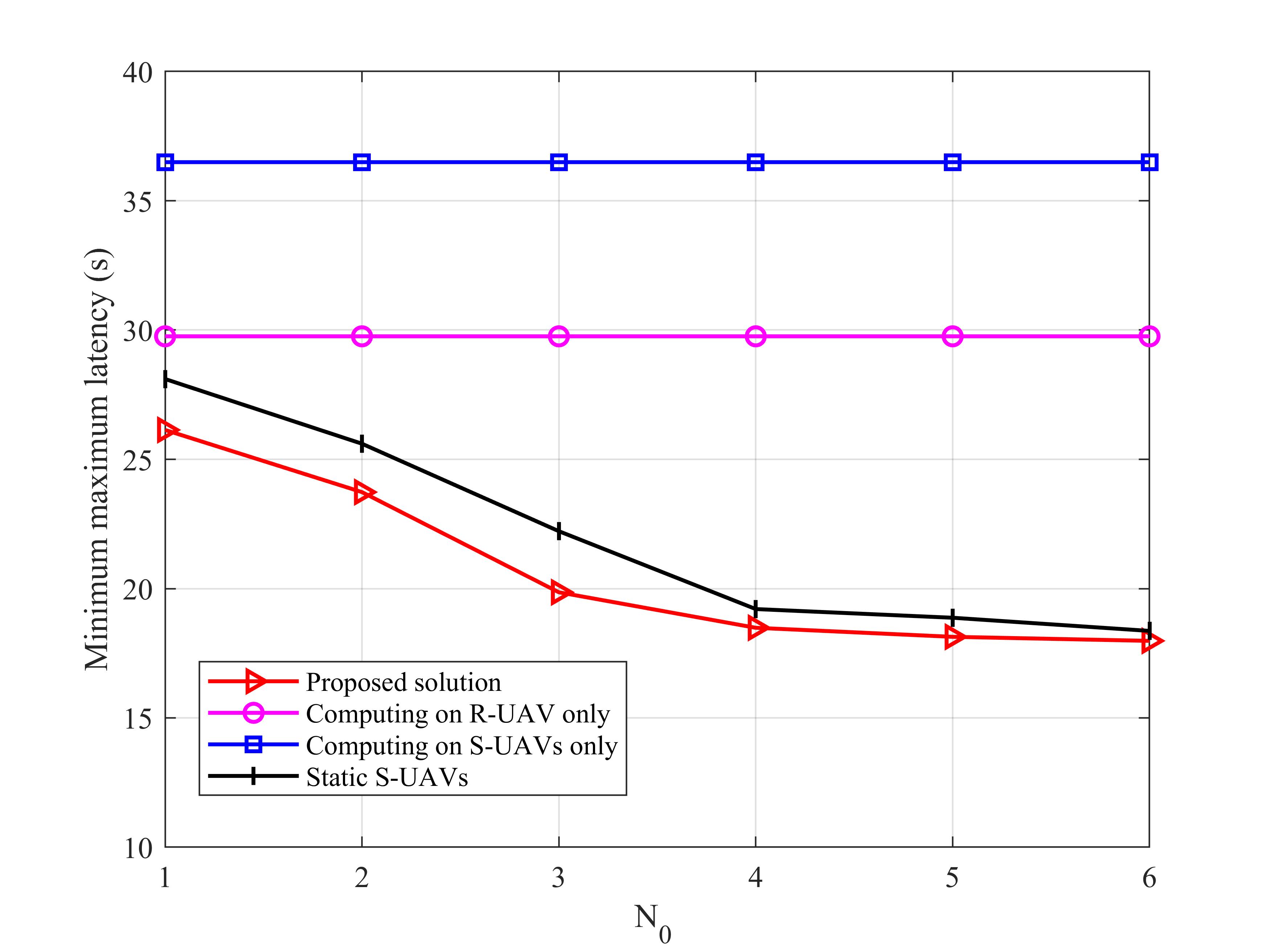}}
\caption{The minimum maximum latency varies with $N_0$.}
\label{Fig.7}
\end{figure}

\begin{figure}[t]
\centering 
{\includegraphics[width=0.45\textwidth]{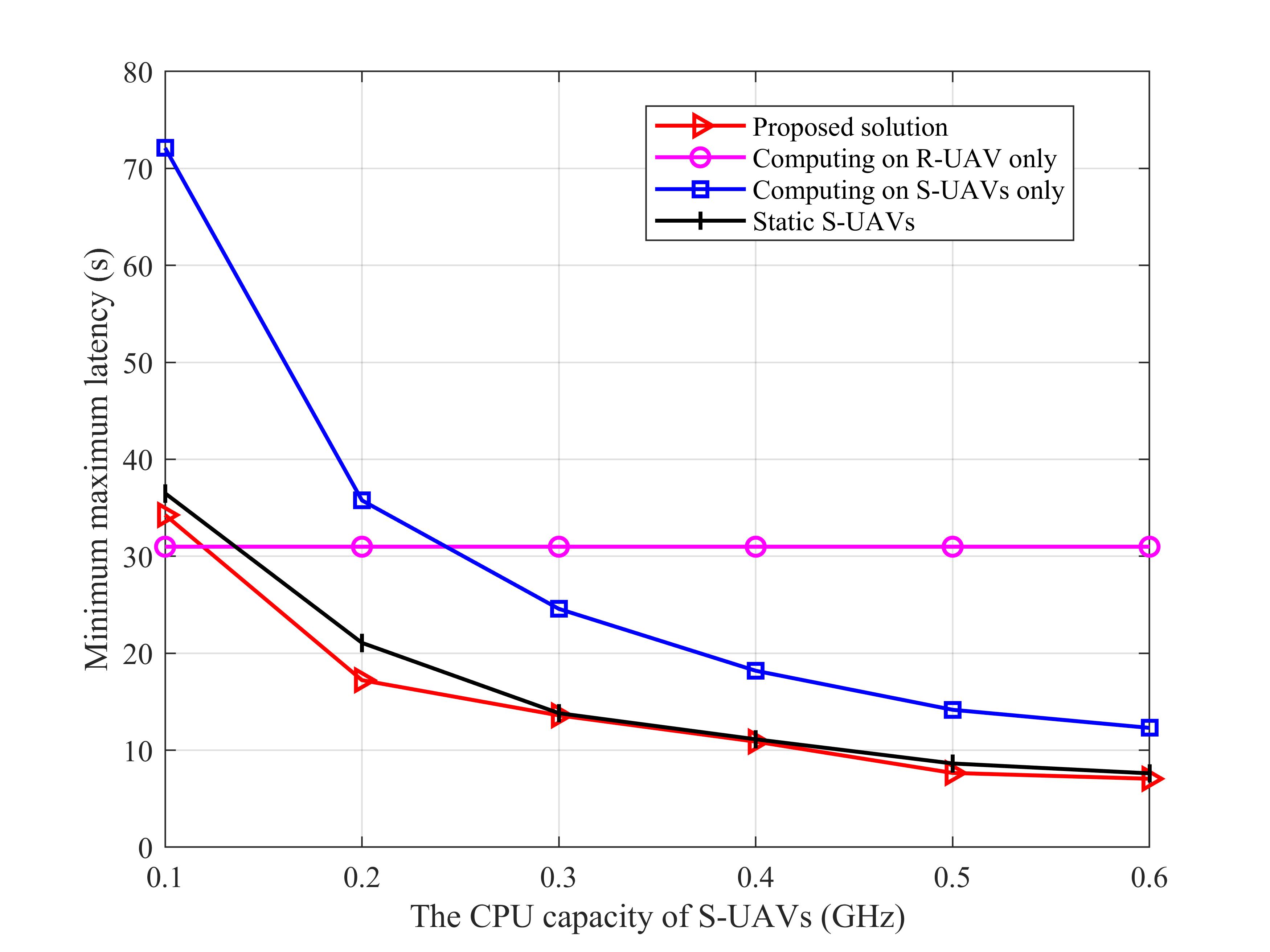}}
\caption{The minimum maximum latency varies with the CPU capacity of S-UAVs.}
\label{Fig.8}
\end{figure}

\subsubsection{Minimum Maximum Latency}
In Fig. \ref{Fig.5}-Fig. \ref{Fig.8}, we evaluate the effects of various factors on the objective value, i.e., the minimum maximum latency. It can be observed that the minimum maximum latency obtained with our proposed solution is significantly lower than Computing on S-UAVs only scheme and Computing on R-UAV only scheme. In addition, the minimum maximum latency obtained by our proposed scheme is always lower than that of the Static S-UAVs scheme, although the gap between them is very small. However, our proposed scheme adjusts the locations of S-UAVs according to the locations of the monitoring targets, which can photograph the rescue targets more closely, so as to obtain a clearer situation.

In Fig. \ref{Fig.5}, we compare the relationship between the minimum maximum latency and the number of video chunks. According to the actual size of the high definition video, the video is divided into multiple video chunks for transmission. As the number of video chunks increases, the objective value of each scheme increases. Moreover, the gap between the Satic S-UAVs and our proposed solution is gradually widening. Unless otherwise specified, the number of video chunks is set to three. In Fig. \ref{Fig.6}, we compare the relationship between the minimum maximum latency and the transmit power of S-UAVs.As the transmit power of S-UAVs increases from 0.2 W to 1 W, the minimum maximum latency of each scheme decreases. Additionally, the gap between each scheme is relatively stable.

\begin{figure}
\centering
        \includegraphics[width=0.45\textwidth]{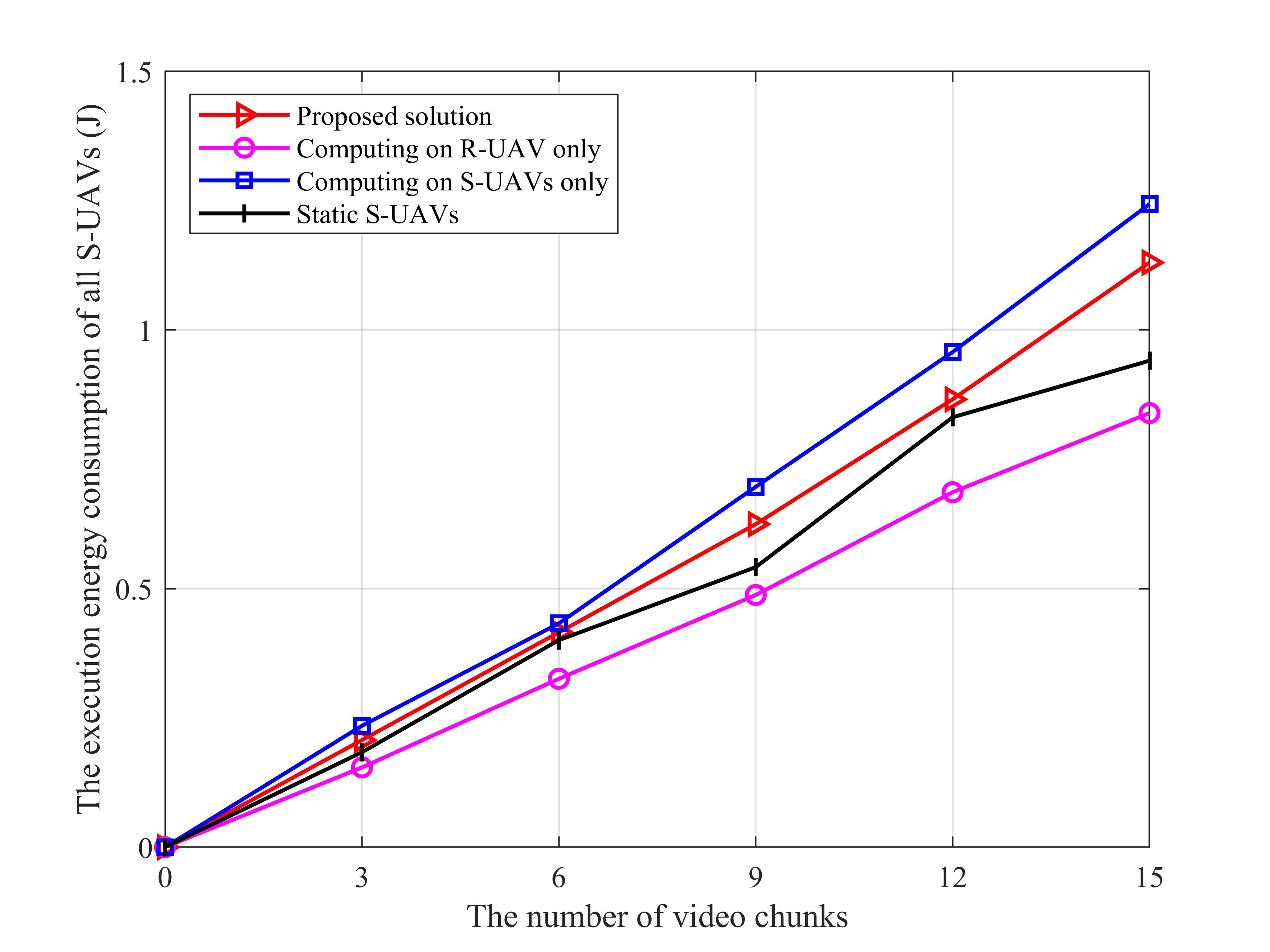}
        \centerline{(a) $p_n=0.2$ W.}
    \hfill
        \includegraphics[width=0.45\textwidth]{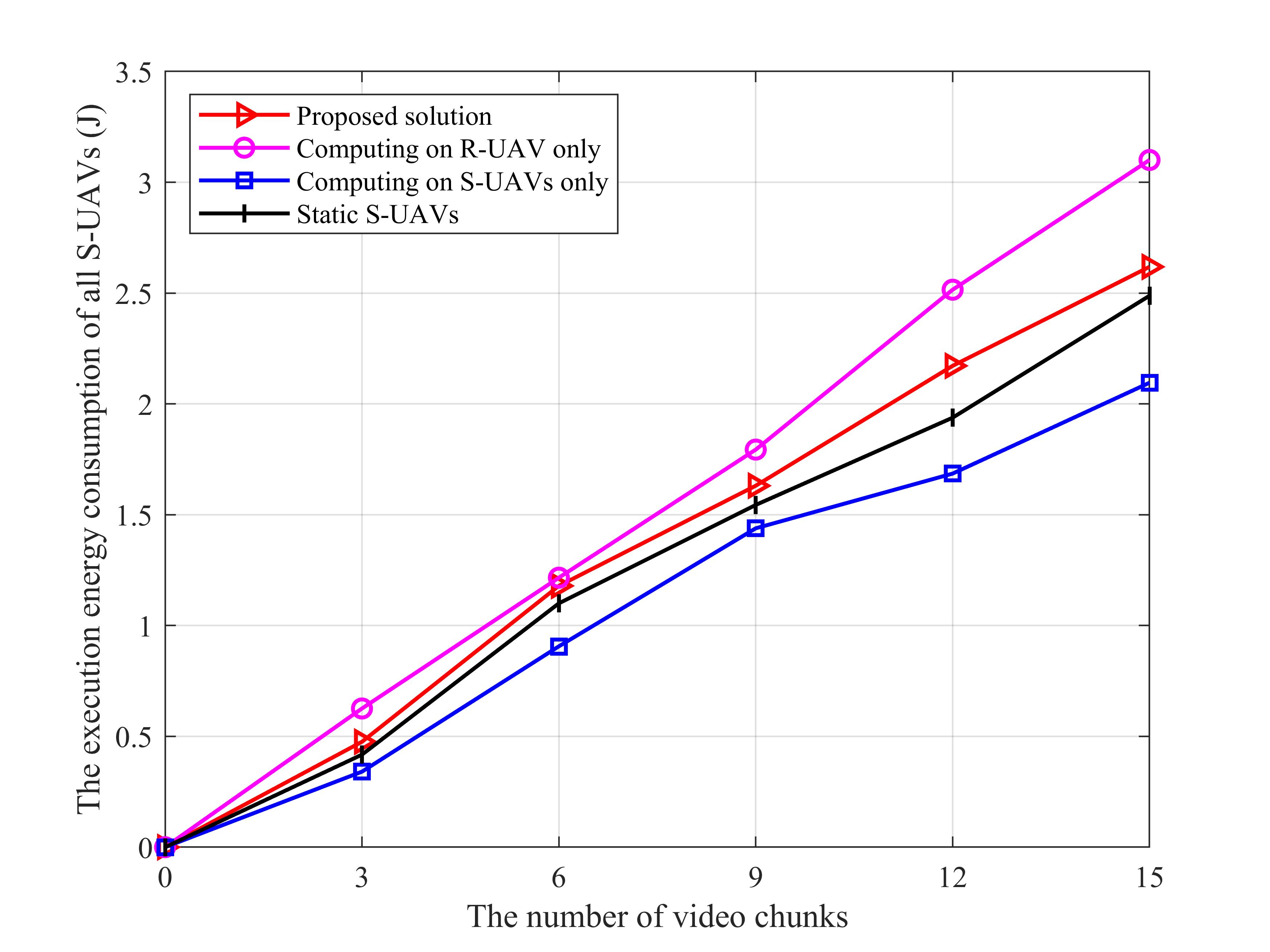}
        \centerline{(b) $p_n=0.8$ W.}
    \caption{The execution energy consumption of all S-UAVs versus the number of video chunks.}
    \label{Fig.9}
\end{figure}

In Fig. \ref{Fig.7}, we compare the minimum maximum latency for different schemes by varying $N_0$. Computing on R-UAV only scheme and Computing on S-UAVs only scheme are independent of changes in $N_0$. As $N_0$ increases, the number of tasks that R-UAV can execute also increases. Since R-UAV has stronger computing capacity than S-UAVs, more tasks are offloaded to R-UAV, resulting in reduced delays. Therefore, the minimum maximum latency of our proposed solution and Static S-UAVs scheme decrease with $N_0$ increases. Additionally, when the value of $N_0$ exceeds four, the variation of the objective value is stable, indicating that when the number of S-UAVs selected to offload the computing task is four, better performance can be obtained.
\begin{figure}[t]
\centering 
{\includegraphics[width=0.45\textwidth]{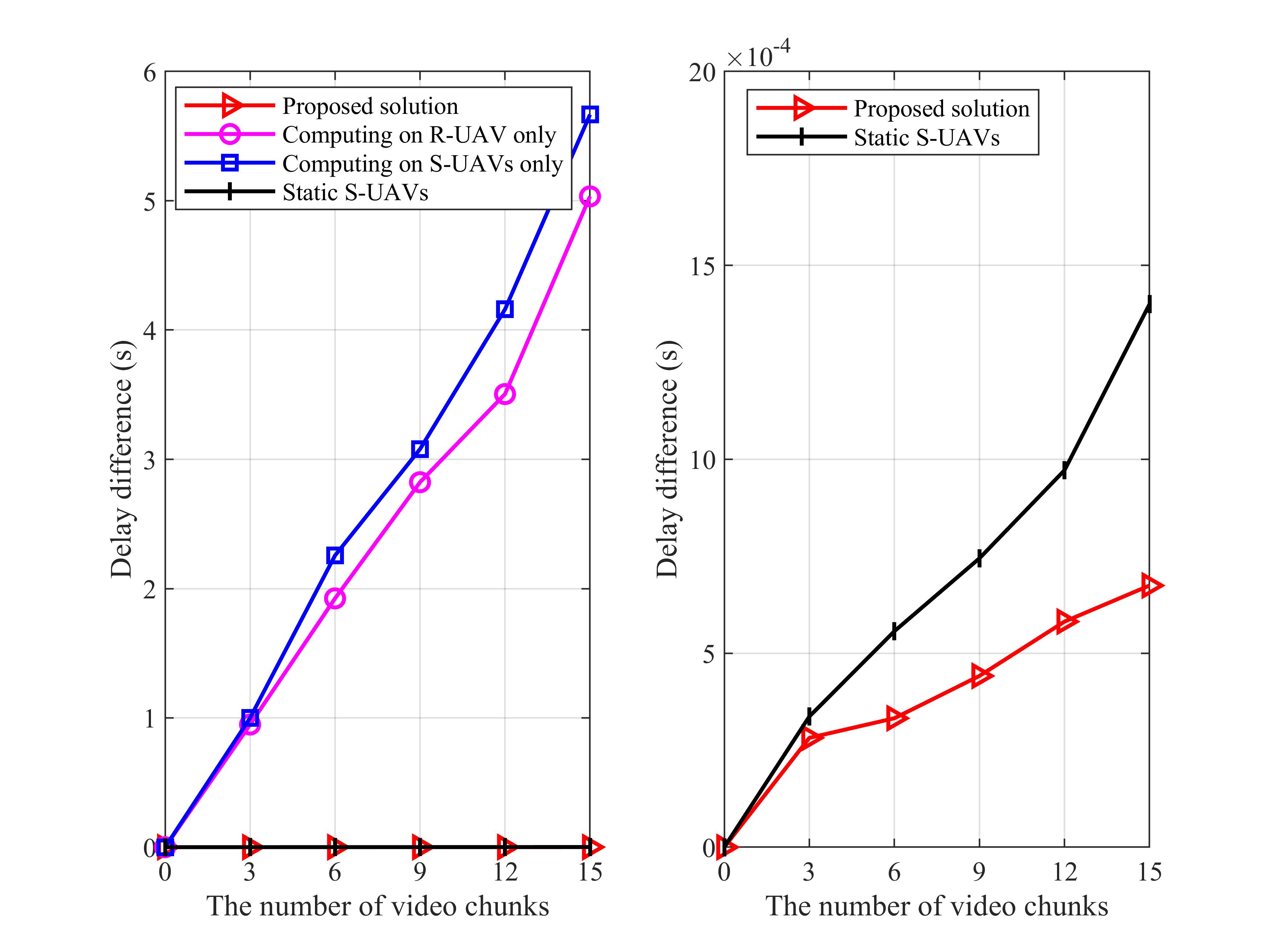}}
\caption{The delay difference versus the number of video chunks.}
\label{Fig.10}
\end{figure}

\begin{figure}[t]
\centering 
{\includegraphics[width=0.45\textwidth]{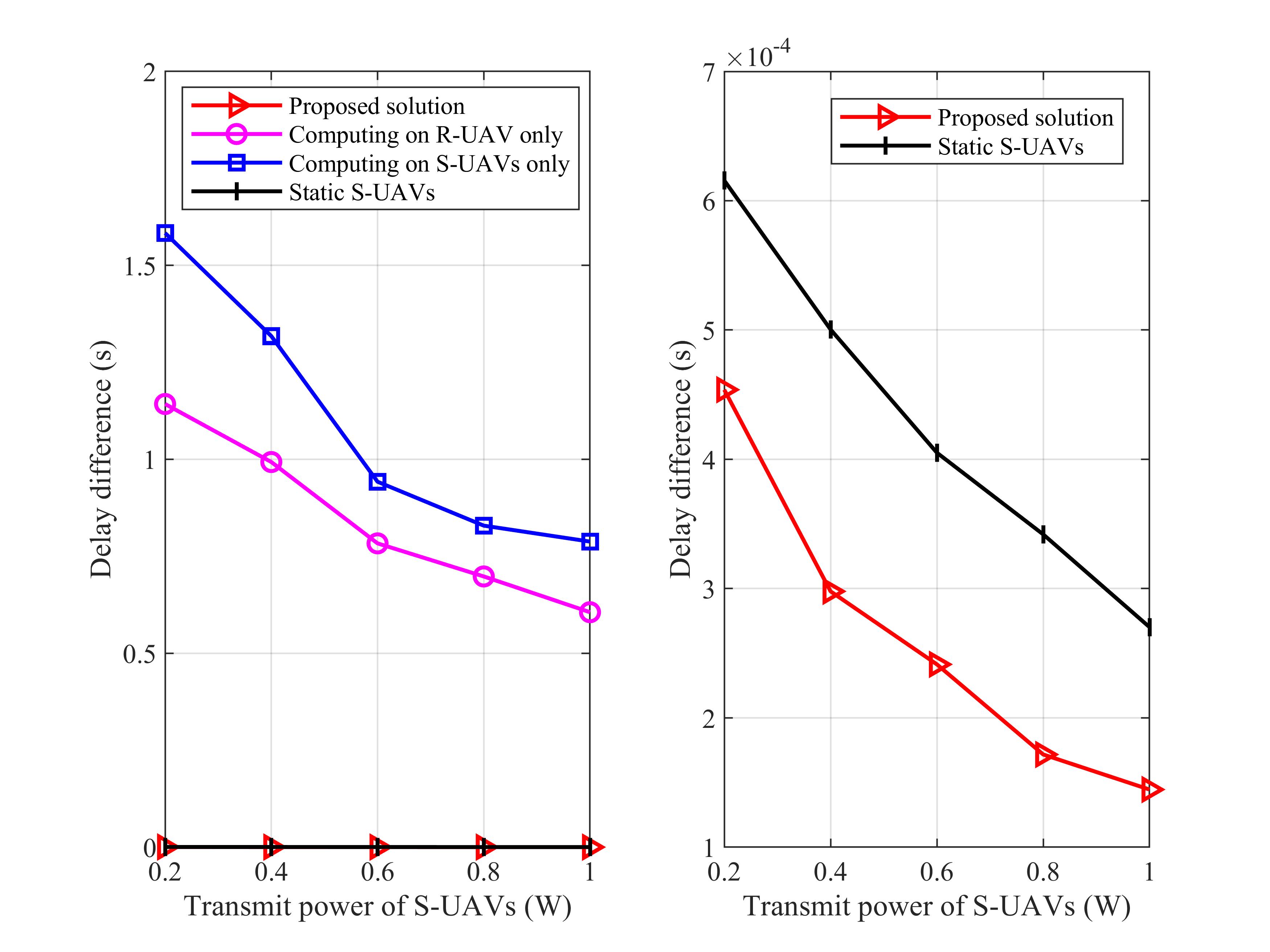}}
\caption{The delay difference versus the transmit power of S-UAVs.}
\label{Fig.11}
\end{figure}

\begin{figure}[t]
\centering 
{\includegraphics[width=0.45\textwidth]{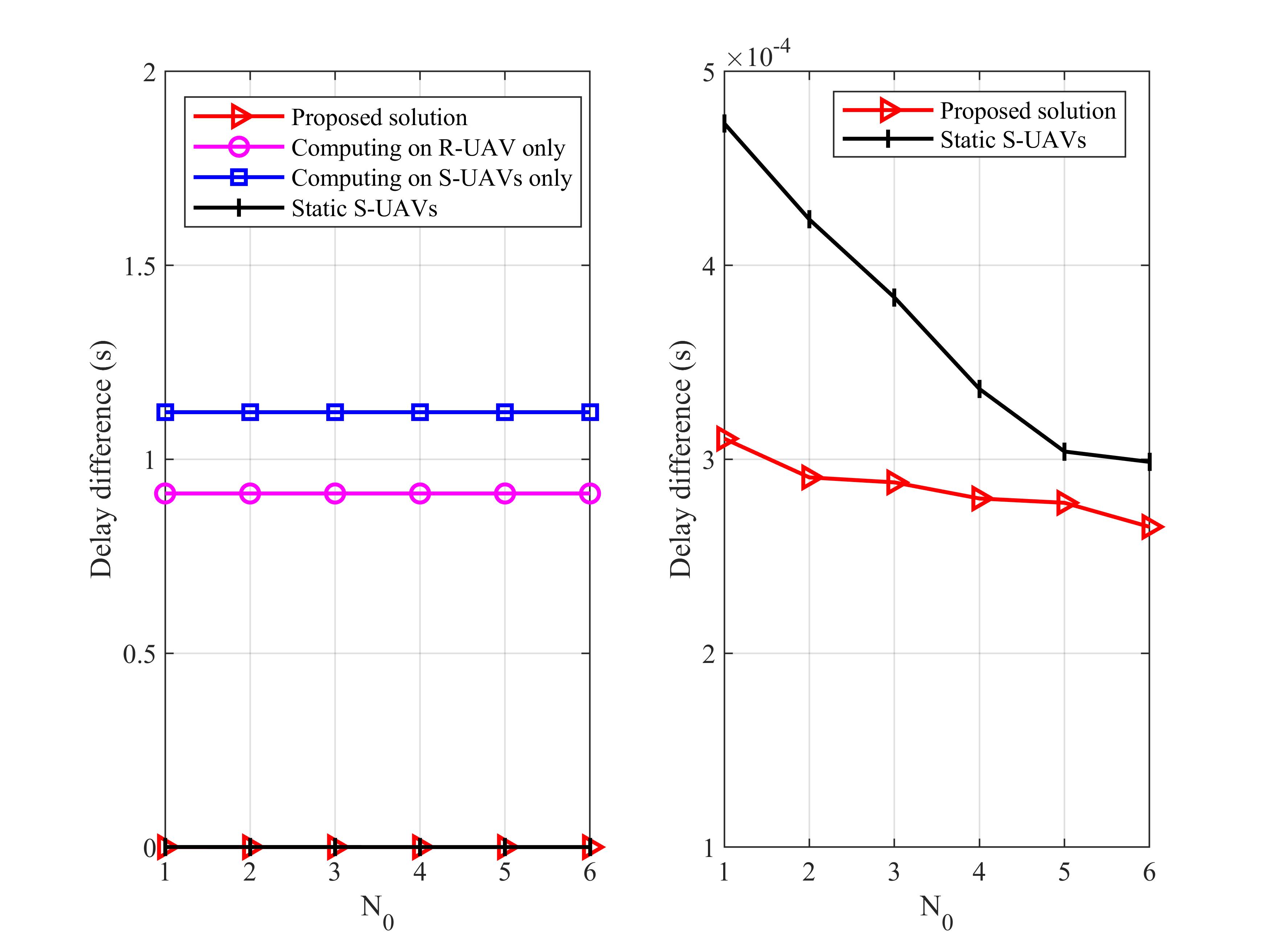}}
\caption{The delay difference varies with $N_0$.}
\label{Fig.12}
\end{figure}

\begin{figure}[t]
\centering 
{\includegraphics[width=0.45\textwidth]{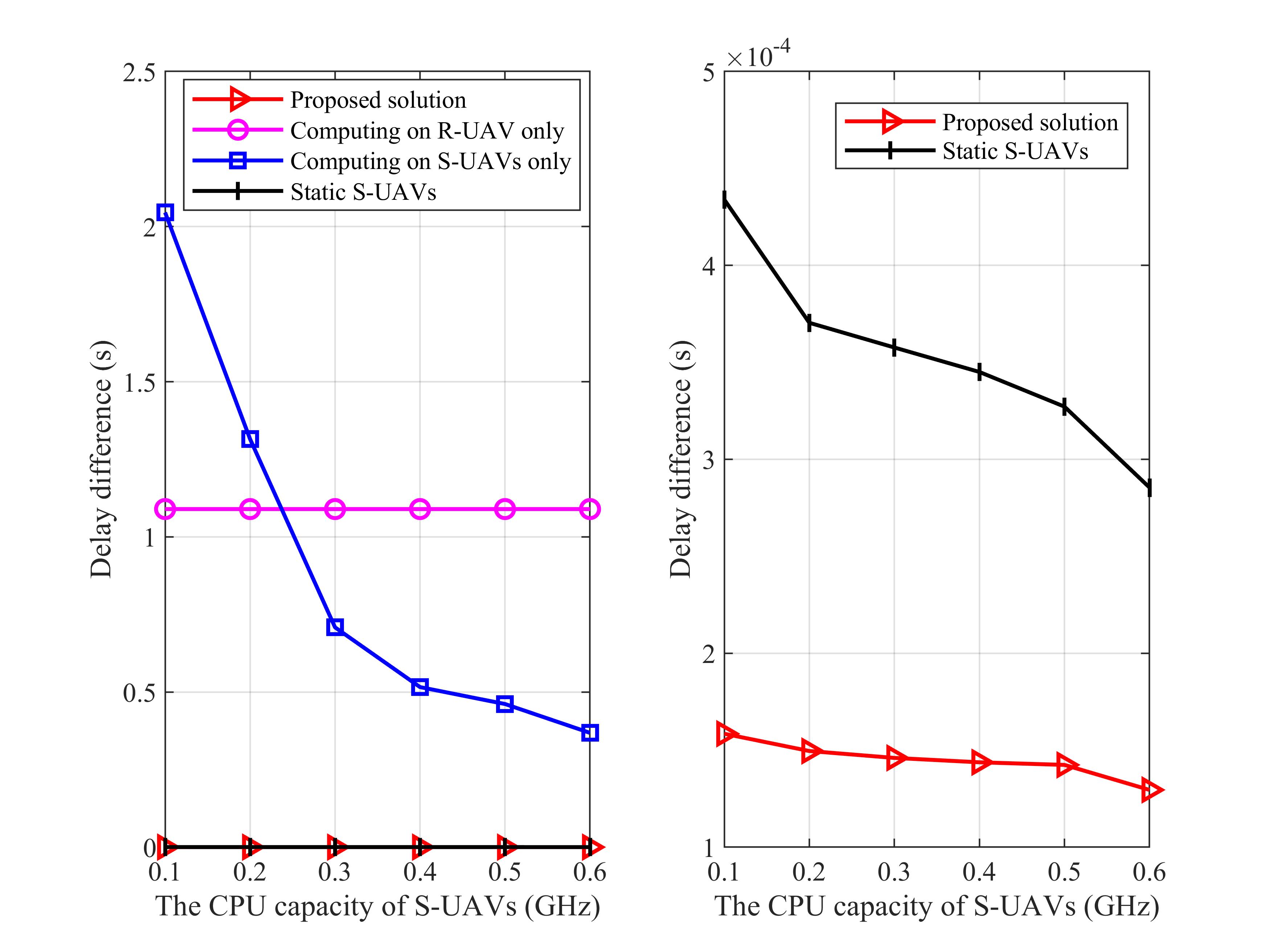}}
\caption{The delay difference varies with the CPU capacity of S-UAVs.}
\label{Fig.13}
\end{figure}
Fig. \ref{Fig.8} shows the trend of the minimum maximum latency as the CPU capacity of S-UAVs increases. Obviously, the change in the CPU capacity of the S-UAV has no effect on Computing on R-UAV only scheme. The objective value decreases as the CPU capacity increases for other three schemes. Additionally, the gap between our proposed solution and the Computing on S-UAVs only scheme decreases with the increase of CPU capacity of S-UAVs. The reason is when the computing power of the S-UAV increases, the computing latency of the S-UAV video processing can be reduced. This allows for a greater inclination towards executing computing tasks locally, resulting in a reduction of data volume and consequently, transmission delay.

\subsubsection{Energy Consumption of S-UAVs}
In Fig. \ref{Fig.9}, we compare the relationship between energy consumption of S-UAVs and the number of video chunks when the transmit power of S-UAVs is 0.2W and 0.8W, respectively. Due to the assumption that the basic operational energy consumption is identical, we only compare the execution energy consumption of S-UAVs, i.e., communication and computation energy consumption. When the transmit power is 0.2W., it can be observed that our proposed solution achieves lower energy consumption compared to the scheme that for Computing on S-UAVs only. Compared with the Computing on R-UAV only and Static S-UAVs scheme, our proposed solution consumes more energy. As the transmit power grows from 0.2W to 0.8W, the energy consumption of all schemes increase. Moreover, the energy consumption of Computing on R-UAVs only is higher then other three schemes when the transmit power is 0.8W. The reason is the increase of the transmit power of S-UAVs lead to the growth of communication energy consumption. However, since latency is of utmost importance in MSAR scenarios, even if energy consumption is not the lowest, it is still within an acceptable range. 

\subsubsection{Delay Difference}
We verify that our proposed solution has a lower latency than the other three benchmarks in the previous analysis. In this part, we aim to make it clear that our proposed solution also has a great effect in reducing the delay difference. The delay difference is defined as the standard deviation of the minimum maximum latency of all S-UAVs. As shown in Fig. \ref{Fig.10}-Fig. \ref{Fig.13}, the delay difference of our proposed solution is notably lower then that of the Computing on S-UAVs only scheme and Computing on R-UAV only scheme. In addition, the delay difference achieved by our proposed solution closely and lower to that of the Static S-UAVs scheme. In addition, the variation trend of delay difference is the same as that of minimum maximum latency.

Fig. \ref{Fig.10} reveals that the delay difference increases with the growth of the number of video chunks. As the number of video chunks increases, the communication delay and computing delay of each S-UAV increase, resulting in increase in the delay difference. Obviously, our proposed solution is more stable. Fig. \ref{Fig.11} shows that the delay difference decreases with the growth of the transmit power of S-UAVs. Each S-UAV's communication delay decreases as transmit power increases, which also reduces the delay difference.

As shown in Fig. \ref{Fig.12}, the setting of $N_0$ has no effect on Computing on S-UAVs only scheme and Computing on R-UAV only scheme. When $N_0$ is changed, the delay difference of Computing on S-UAVs only scheme and Computing on R-UAV only scheme remain unchanged. The delay difference of our proposed solution and Static S-UAVs decreases with the increase of $N_0$. Similarly, the CPU capacity of S-UAV has no effect on Computing on R-UAV only scheme, when the CPU capacity of S-UAV changes, the delay difference of Computing on R-UAV only scheme remain unchanged. When the CPU capacity of S-UAV improves, the delay difference of the other three schemes reduce, and the delay difference obtained by our proposed solution is the smallest, as shown in Fig. \ref{Fig.13}.

\section{Conclusion}
In this paper, we have considered the problem of minimizing the maximum total latency among all S-UAVs for the multi-UAV assisted MSAR system. Our approach involves optimizing the position of R-UAV, the association between S-UAVs and rescue targets, and the offloading decisions to achieve this goal. Unlike existing works, we have taken into account the S-UAV position adjustment process to achieve the monitoring integrity and fairness for rescue targets. Since the formulated problem is an MINLP problem, we have decomposed it into three sub-problems and developed an overall iterative algorithm to find the sub-optimal solutions. We have conducted extensive numerical simulations to validate the effectiveness of the proposed algorithm. The numerical results have demonstrated that our proposed solution achieves a lower latency and a smaller delay difference compared to baseline schemes. For future work, the energy efficiency optimization scheme for MSAR systems can be further investigated by considering the limited energy of UAVs. Besides, the multi-hop transmission scheme for MSAR systems can be further considered.

\bibliographystyle{IEEEtran}
\bibliography{ref}

% \vfill

\end{document}